\documentclass[aps,pre,twocolumn,showpacs,floatfix,superscriptaddress]{revtex4-1}
\usepackage{hyperref,graphicx}	
\usepackage{natbib}
\usepackage{array}
\usepackage{inputenc}

\newcommand{\Bo}{\textrm{Bo}}
\newcommand{\xmax}{x_\textrm{\tiny max}}
\newcommand{\xmin}{x_\textrm{\tiny min}}
\hypersetup{colorlinks,%
   	citecolor=[rgb]{0.6,0,0},%
   	filecolor=[rgb]{0,0.6,0},%
   	linkcolor=[rgb]{0,0,0.5},%
   	urlcolor=[rgb]{0,0,0.5},%
  	}
\begin{document}

\title{Model for Polygonal Hydraulic Jumps}
\author{Erik A. Martens}
\affiliation{
Group for Biophysics and Evolutionary Dynamics, MPI for Dynamics and Self-Organization, 37073 G\"{o}ttingen, Germany}
\email{erik.martens@ds.mpg.de}
\altaffiliation{Theoretical and Applied Mechanics, Cornell University, Ithaca, NY 14853, USA}
\altaffiliation{
Center of Fluid Dynamics and Department of Physics, Technical University of Denmark, 2800 Kgs.~Lyngby, Denmark}
\author{Shinya Watanabe}
\affiliation{Department of Mathematics and Informatics, Ibaraki University, 310-8512, Mito, Japan}
\author{Tomas Bohr}
\affiliation{Center of Fluid Dynamics and Department of Physics, Technical University of Denmark, 2800 Kgs.~Lyngby, Denmark}


\begin{abstract}
We propose a phenomenological model for the polygonal hydraulic jumps discovered by Ellegaard {\it et al.}, based on the known flow structure for the type II hydraulic jumps with a ``roller'' (separation eddy) near the free surface in the jump region. The model consists of mass conservation and radial force balance between hydrostatic pressure and viscous stresses on the roller surface. In addition, we consider the azimuthal force balance, primarily between pressure and viscosity, but also including non-hydrostatic pressure contributions from surface tension in light of recent observations by Bush {\it et al}. The model can be analyzed by linearization around the circular state, resulting in a parameter relationship for nearly circular polygonal states. A truncated, but fully nonlinear version of the model can be solved analytically. This simpler model gives rise to polygonal shapes that are very similar to those
observed in experiments, even though surface tension is neglected, and the condition for the existence of a polygon with N corners depends only on a single dimensionless number $\phi$. Finally, we include time-dependent terms in the model and study linear stability of the circular state. Instability occurs for sufficiently small Bond number and the most unstable wave length is expected to be roughly
proportional to the width of the roller as in the Rayleigh-Plateau instability.  



\begin{center}
 {\it }
\end{center}
\end{abstract}
\maketitle



\section{Introduction}
\begin{figure}[ht!]
  \includegraphics[width=0.4\textwidth]{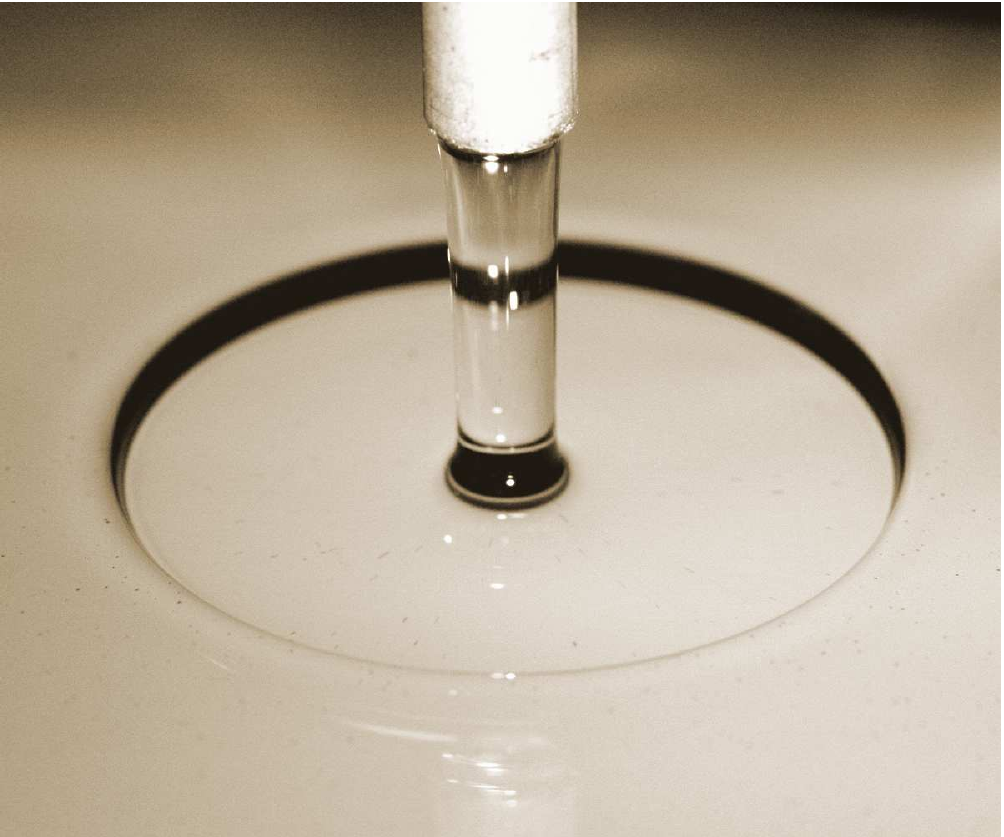}
  \caption{Circular hydraulic jump. A cylindrical jet of fluid impinges vertically on a horizontal glass plate  and forms a circular hydraulic jump. The fluid used here is ethylene glycol (about 10 times the viscosity of water).}
\label{fig:circularjump}
\end{figure}

\begin{figure}[ht!]
  \includegraphics[width=0.475\textwidth]{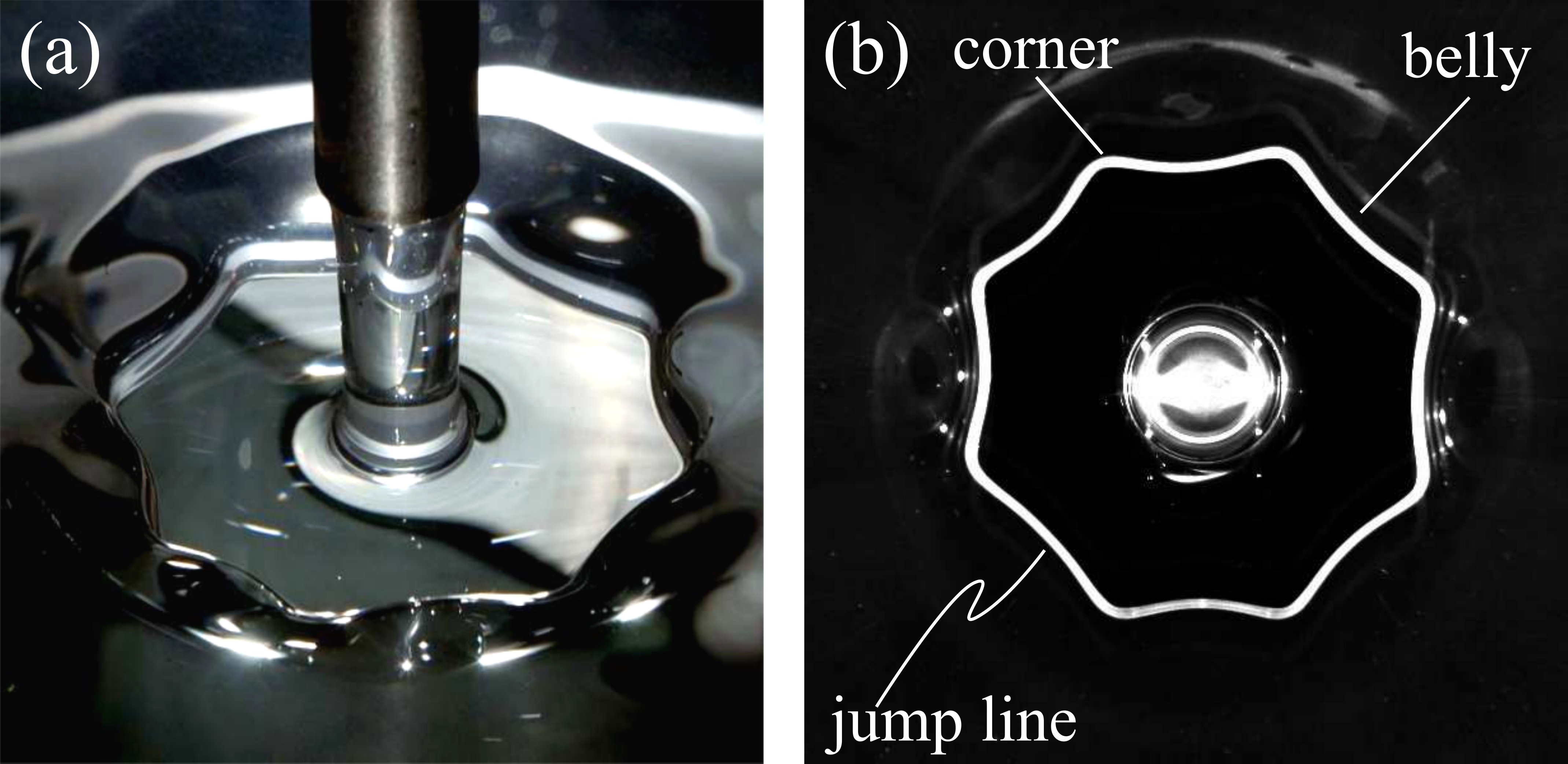}\
  \caption{ Polygonal jump with eight corners. {\bf (a):} Top view from an oblique angle. {\bf (b):} View from below through a glass plate. The jump line becomes visible by diffracting light shone  on the jump from above. The  ``belly'' between two corners (or ``necks'') is where the roller is thickest.}
\label{fig:polyjump}
\end{figure}

The term ``hydraulic jump" refers to the sudden jump in fluid height, as for example observed in the outward spreading water layer in a kitchen sink, resulting from the impact of the water from the tap with the horizontal bottom of the sink \cite{Rayleigh, Watson, Olsson, Chow}. Similar phenomena are seen in rivers with large tidal variation at the outlet and are known as ``river bores". River bores move up the rivers, whereas hydraulic jumps are stationary, either due to spatial inhomogeneities or to the geometric configuration of the flow. In the kitchen sink, the jump occurs on a more or less circular locus. Close to the point of impact the water level is very thin, but at a certain radius, $r_j (\theta)$, the level increases abruptly forming a circular jump. The water flow and thus the jump shape in a kitchen sink fluctuate, but by building a more symmetric experimental setup and/or using a more viscous fluid, one may obtain a completely stationary and axisymmetric flow, where the jump occurs on a surprisingly well-defined circle  (FIG.~\ref{fig:circularjump}). It has been shown  \cite{Nature,Nonlinearity,PhysicaB} that circular hydraulic jumps can undergo a sequence of structural changes seen by varying the fluid height $h_o$ downstream of the jump. 
This can be achieved by inserting an adjustable circular weir at the rim of the circular impact plate. When increasing $h_o$, the jump becomes steeper until, at a critical value of $h_o$, the jump becomes unstable and loses its balance.
Like a breaking wave, it creates a new stationary state with a wider jump region and a new flow structure in which the surface flow is
reversed due to a separation vortex (referred to in the following as the ``roller''). Following \cite{PhysicaB}, his new flow structure is called a {\it type II jump} to distinguish it from the ``ordinary'' {\it type I jump}. The most remarkable observation about the type II state is that the jump typically loses its azimuthal symmetry and attains the shape of a regular, though not necessarily straight-edged, polygon (FIG.~\ref{fig:polyjump}). 

In this paper, we present a phenomenological model for the type II jump, which we believe to include the basic mechanisms of the flow rendering the circular state unstable and giving the polygonal jump its peculiar shape.  This allows us to obtain the qualitative dependence of the jump shape on the control parameters available, primarily the outer fluid height $h_o$, and the properties of the fluid, i.e., viscosity and surface tension.  For experimental data, we refer to earlier work. Our own experiments are performed with the same setup as in \cite{Nonlinearity} and is only used qualitatively as a visual guide (see also supplementary movies~\cite{suppmaterial}).

For the type I state, averaging theory with a variable profile gives at least qualitatively the right flow structure including the separation region on the bottom, outside the jump \cite{Bohr, WPB}. For the type II state with separation at the surface, no such theory exists, and we therefore from the outset assume a flow structure derived from experiments.  
Our model extends earlier models proposed in  \cite{Nature,Nonlinearity,RUC}, based on the competition between gravitational and viscous effects. In \cite{Nature,Nonlinearity} a radial force balance for the roller was used together with a ``line tension" to derive polygonal shapes. In \cite{RUC} it was shown that such a line tension is not necessary and that a more satisfactory model is obtained by taking into account the azimuthal flow. This is the approach used in the present paper. It was recently pointed out \cite{Bush}, however, that the instability forming the polygonal jumps seems to be driven, at least in part, by surface tension (see also Supplementary Materials~\cite{suppmaterial}). Thus, surface tension is included in our model and indeed the instability forming the polygonal jumps seems to be closely related to the Rayleigh-Plateau instability for the ``liquid cylinder" formed by the roller.

In this paper, we treat both linear and nonlinear properties of our model. For the linear properties, we can with some confidence include the effects of surface tension, but for the fully developed polygonal states, this is very difficult because of the lack of precise data on the height profiles. Even without surface tension, our model does have polygonal states, and  we compute their shapes  assuming that they will be only slightly changed by surface tension effects.
The layout of the paper is as follows: we first derive the model and point out the similarities and differences to earlier models. We then solve a nonlinear model with zero surface tension analytically and show that it has many features that are expected from experiments. Finally, we compute the temporal stability of the circular state while including surface tension, and show that it has properties close to the Rayleigh-Plateau instability.

\section{Derivation of the model}\label{sec:model_derivation}
\begin{figure}
 \includegraphics[width = .4\textwidth]{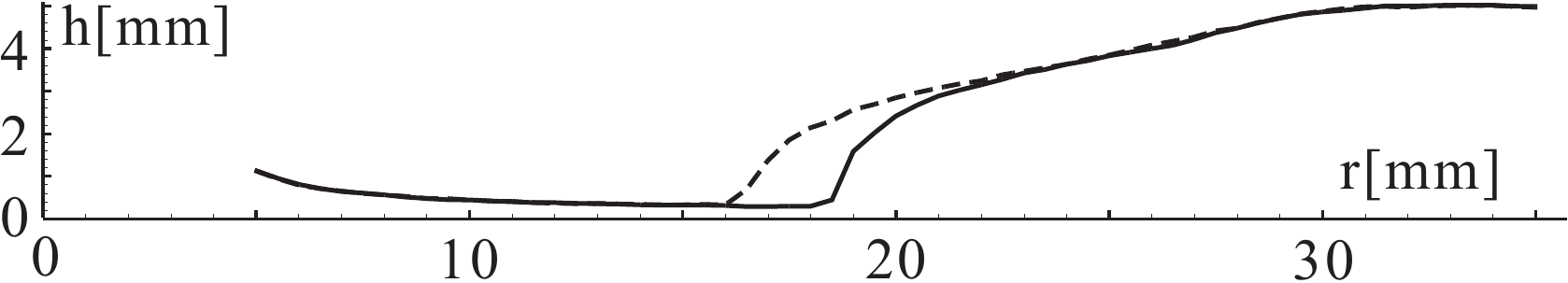}
  \caption{Measured height profiles of a pentagonal polygonal hydraulic jump in ethylene glycol. The solid curve represents the profile measured along the radial direction through a corner. The dashed curve represents the profile along the radial direction midway between two corners (''belly''), where the edge of the jump is almost straight (see e.g. Fig.~\ref{fig:polyjump}(b)). The height of the fluid layer was measured manually with a depth micrometer attached to a translation table. Reproduced with permission from \cite{1999ma}.
  The flow rate is $Q = 40 \text { ml s}^{-1}$. 
  \label{fig:heightprofile}}
\end{figure}

In the experiment \cite{Nature,Nonlinearity,PhysicaB}, the fluid falls from a nozzle mounted at some height, at constant volumetric flow rate, $Q$. We employ cylindrical coordinates, $(r,\theta,z)$, centered at the point, where the central axis of the cylindrical liquid jet impinges the horizontal plate. 
The height of the fluid surface above the plate is parametrized by $h(r,\theta)$. As the fluid leaves the point of impact, $h$ is small and the fluid spreads in a very thin layer (around 1 mm in our experiments) with supercritical flow speed (i.e.~a speed larger than the speed of
surface waves). 
At a short distance from the point of impact, the boundary layer in the film has become fully developed \cite{Watson} and the fluid height is almost constant, i.e., $h(r<r_j,\theta) \approx h_i$, as seen in FIG.~\ref{fig:heightprofile}.
Due to the supercriticality, this inner flow is not affected by the transition from type I to type II jump, and the flow remains axially symmetric.
The surface height of the type II jump abruptly increases at the jump radius $r_j$ due to the presence of the roller on top of the thin fluid layer. 
The surface height increases for $r_j<r$ until a certain radius $R$ where it settles at a roughly constant level, $h(r>R,\theta) \approx h_o$. 
The height difference across the jump is $\Delta h$, and since $h_o \gg h_i$, we use the approximation
\begin{equation}
  \Delta h \equiv h_o - h_i \approx h_o
\label{eqn:deltah}
\end{equation}
throughout the paper. We further introduce the {\em aspect ratio}
\begin{equation}
\label{alpha}
\alpha = {\frac{h_o}{R}}
\end{equation}
which is typically small, say of the order of 0.2.

\begin{figure}
  \includegraphics[width = .45\textwidth]{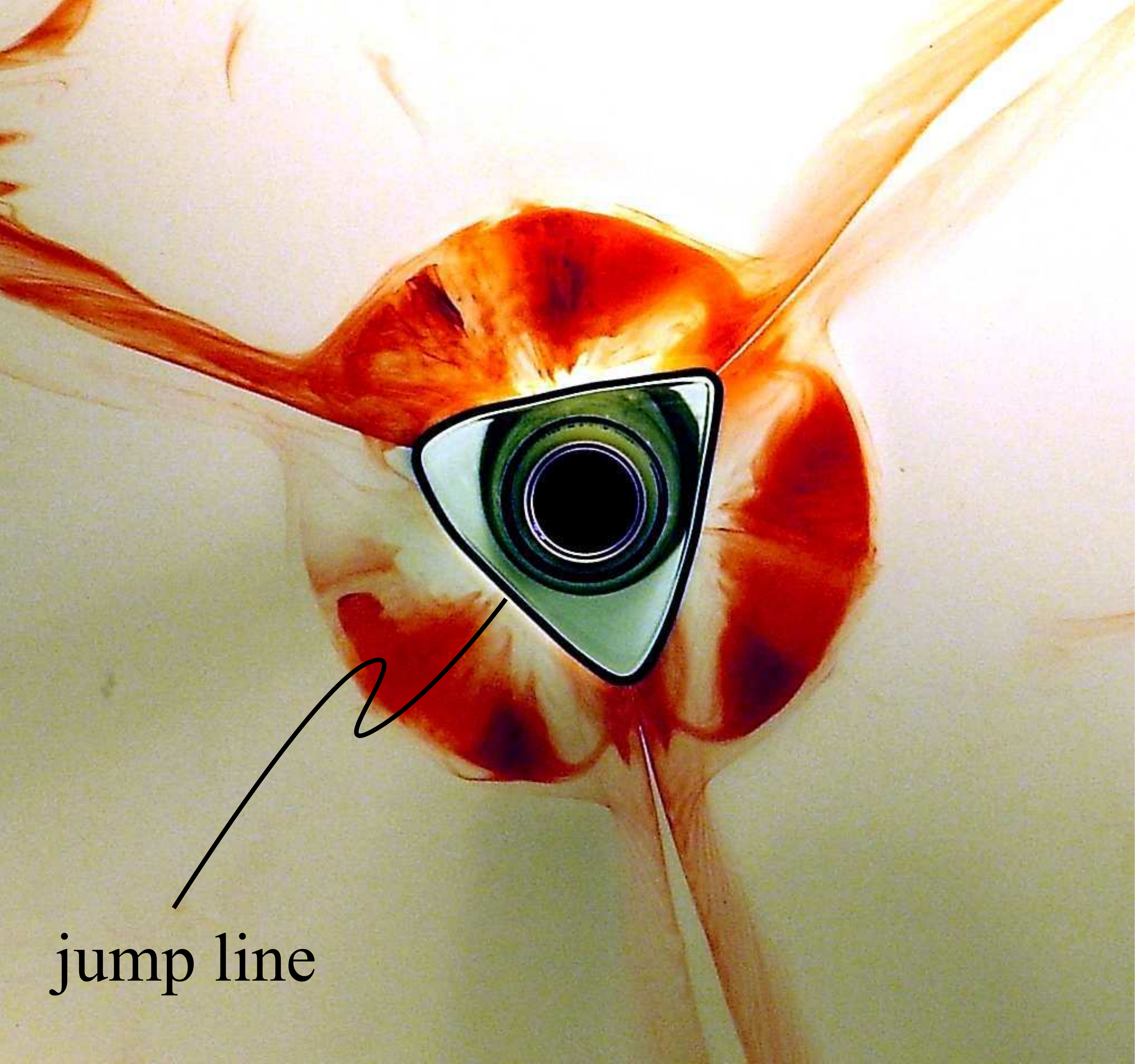}
  \caption{(Color online) Flow visualization of the roller vortex in the case of a triangular jump in ethylene glycol. The jump is seen from below through a glass plate, where the impinging jet is visible as the black center region and the jump line as the black line surrounding it. Red dye is injected into the roller flow by letting droplets of the dye fall from above into the vortex. It is seen that the roller structure extends from the jump line to the outer radius. In the corner region, fluid is expelled in clearly visible radial jets. Note also the fine white line between the roller vortices, which indicate that the vortices are actually disconnected structures. For a visualization of the flow see also Supplementary Material~\cite{suppmaterial}.
  \label{fig:rollerflowvisual}}
\end{figure}

In FIG.~\ref{fig:rollerflowvisual} we show a visualization of the flow in
a triangular jump created with ethylene glycol. It can be clearly seen
that the roller fills the jump region from $r_j$ to $R$, where $r_j$ now
has lost its azimuthal symmetry, whereas the outer edge $R$ remains
circular. We therefore define the
normalized local roller width as one of the basic variables of our model  
\begin{equation}
   \delta(\theta) = \frac{R-r_j(\theta)}R  \qquad ( 0 < \delta < 1 ) ,
\end{equation}
which varies with $\theta$ to give the jump its characteristic polygon
shapes. The dye dripped into the roller reveals an extremely slow exchange
of the fluid in the roller (see also flow visualization in the Supplementary Material~\cite{suppmaterial}). In fact the dye can be visible in the roller for several minutes if it is left undisturbed. From this we conclude that the main part of the fluid in the thin layer going under the roller leaves
the jump region without ever entering the roller. Downstream of the jump region, 
the flow speed is subcritical (less than the speed of surface waves), and the flow is
again purely radial, but it now has an azimuthal dependence: at the corners
of the polygon, strong radial jets are observed. Within the jump region, an azimuthal transport must therefore exist in the jump region from the sides to the corners of
the polygon. Indeed a small, spiraling azimuthal flow is observed in the
roller from its ``belly" out to the ``neck''.
From FIG.~\ref{fig:rollerflowvisual} the jets seem to consist primarily of liquid
going right through the jump, but they also carry with them the dyed
fluid transported to the corners inside the roller. The jets at
the corners of an $N$-gon are actually so strong that they apparently break up the
roller into $N$ distinct rollers, touching and interacting in a rather
complicated manner in the corners \cite{Nonlinearity}. This
effect is not included in our model, but since it occurs only in narrow
regions near the corners, we believe that our model can still give useful
results.
\begin{figure}
  \includegraphics[width=0.45\textwidth]{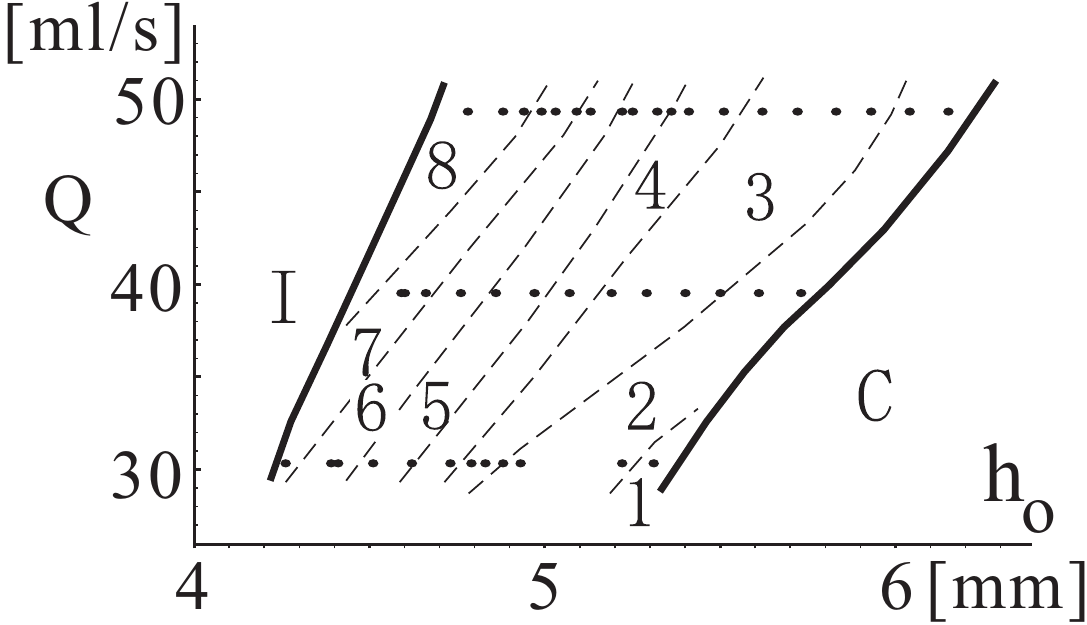}\\
\caption[Measured phase diagram]{
Measured phase diagram, reproduced with permission from \cite{Nonlinearity}.
Polygonal hydraulic jumps are observed using ethylene-glycol, which has a viscosity about 10 times that of water.
Polygons appear in the parameter regime between the circular type-I state (marked as `I') and the closed state (i.e. no jump, marked as `C')
when the jet flux $Q$ and the outer height $h_o$ are varied. The height of the nozzle about the surface is set to 2 cm.
The number of corners is found to be less sensitive to the nozzle height \cite{Nonlinearity}.
Polygonal states with the number of corners $N=1,2,\dots,8$ are found in this experiment.
\label{fig:phasediagram} }
\end{figure}
A phase diagram has been presented in \cite{Nature} in terms of
three parameters: the flow rate $Q$, the outer fluid height $h_{o}$
and the height of the nozzle.
The dependence on the first two parameters
is reproduced in FIG.~\ref{fig:phasediagram}.
When increasing $h_o$, the transition to the type
II structure leads to a polygon with many corners (up to $N=13$ have been
observed). Increasing $h_o$ further, the number of corners decreases
and so does the average jump radius, $(2 \pi)^{-1}\oint
r_j(\theta)d\theta$, until at last the jump closes on the vertical jet
from the nozzle. Throughout this regime the ''outer'' radius $R$ is surprisingly constant, 
while the inner border defining the polygon changes. 
In fact, this polygon shape shows considerable 
hysteresis, and a whole series of bordering polygon shapes can be stable at the same flow rates.
An extra corner can be created by pulling out a polygon side with a needle, and similarly, a corner
can be removed by pushing a corner inward. For sufficiently small
disturbances the jump simply deforms - just as if a rubber band was pulled
- and bounces back to its original shape after being released.

\subsection{Momentum Balance for the Roller}
\label{sec:controlvolume}
\begin{figure}
    \includegraphics[width = .475\textwidth]{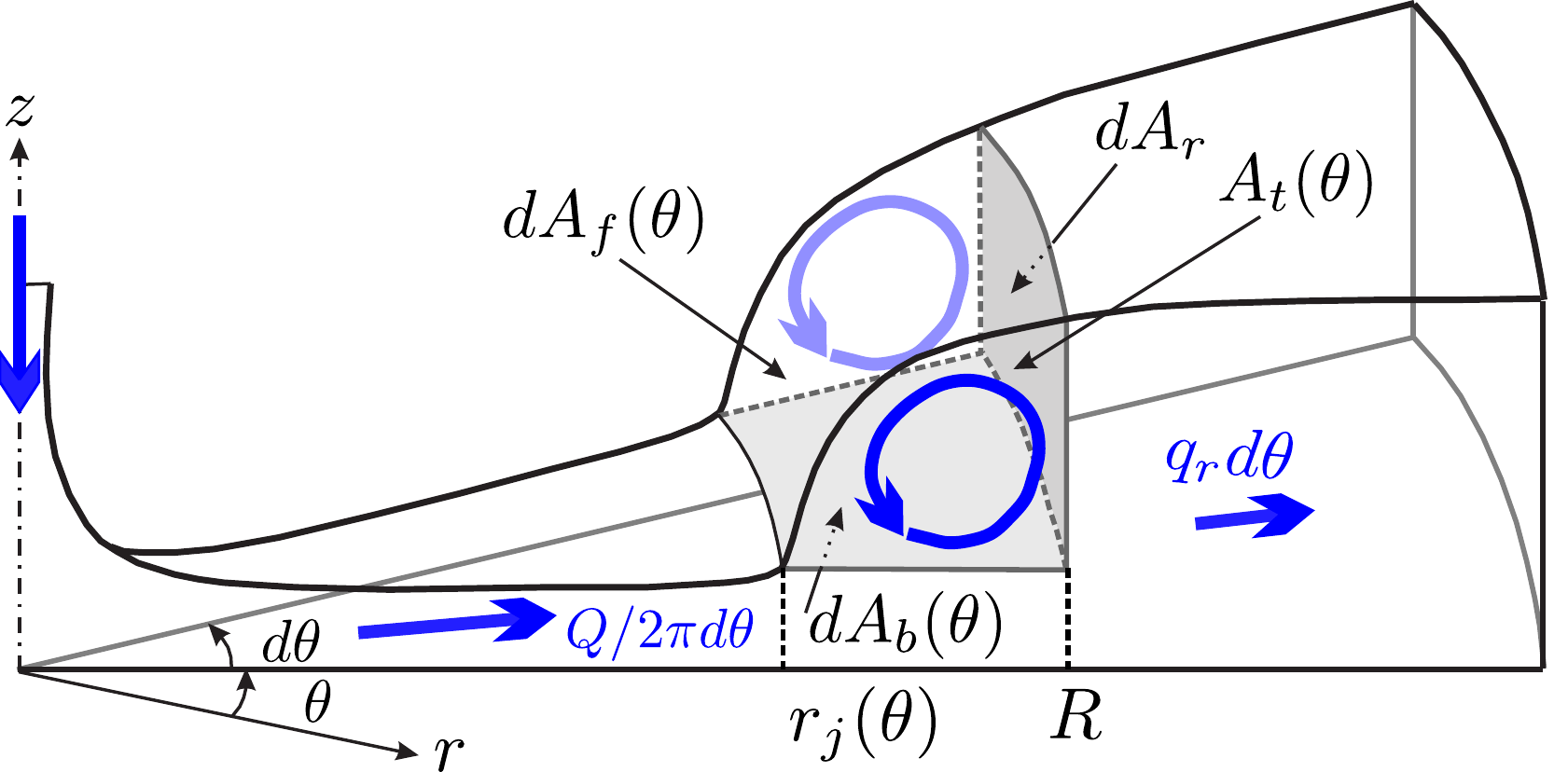}
\caption{(Color online) 
A three-dimensional view of the infinitesimal wedge $[\theta,\theta+d\theta]$
for the hydraulic jump. The gray shaded region is a part of the roller eddy near the surface
just after the jump, and represents the control volume we consider
for the mass and force balances. Areas of the side, bottom, rear faces, as well as the free-surface portion of the volume
are labeled as $A_t(\theta)$, $dA_b(\theta)$, $dA_r$, and $dA_f(\theta)$, respectively. The bold blue arrows indicate the flow direction.
\label{fig:jumpanatomy}}
\end{figure}

In our phenomenological model,  we focus on the roller which we consider as a separate object interacting with the fluid beneath and behind it. We look at a control volume defined by the section of the roller inside an infinitesimal wedge $[\theta, \theta + d\theta]$, as illustrated in FIG.~\ref{fig:jumpanatomy},
and label the areas of the bottom and the rear faces of the control volume
as $dA_b(\theta)$ and $dA_r$, respectively.
The lateral areas are denoted as $A_t (\theta)$.
Let further $\tau_{ij}$ be the stress tensor, let
$V$ and $S$ be the volume and the boundary, respectively, of the roller slice,
and let $\mathbf{\hat n}$ be the unit normal of $S$.
Then the momentum equations are
\begin{eqnarray}\label{eq:momentumbalance}
   && \int_V  \frac{\partial}{\partial t}\left(\rho \mathbf{u}\right)dV +\int_S \rho\mathbf{u} (\mathbf{u}\cdot\mathbf{\hat n}) dS  \nonumber \\
    &&= -\int_S  p\, (\mathbf{\hat n}\,dS) +  \int_S \tau_{ij}
    (\mathbf{\hat n}_jdS) + \int_V \rho \,\mathbf{f} dV.\
\end{eqnarray}
Initially, we shall consider stationary states and thus the first term on the left hand side is zero. Also, the last term on the right vanishes, since the only body force is gravity, which is taken into account by assuming that the pressure $p$ is hydrostatic. 

The inner flow in FIG.~\ref{fig:jumpanatomy} is purely radial and axially symmetric. When it reaches the roller, most of it passes underneath it. When it has passed the roller the flow is still radial, but has acquired an angular dependence. In between, in the region of the roller, the flow thus acquires a small tangential component, which we assume to run inside the roller. Our first task is then to write down the continuity equation for the flow. Then we shall evaluate the ``force" terms in (\ref{eq:momentumbalance}) for the radial and tangential directions, respectively.
In the following, we use cylindrical coordinates and denote the respective vector components of dependent variables by indices $r, \theta$ and $z$.

\subsection{Mass Conservation}
\label{sec:massconservation}
\begin{figure}[tbh]
    \includegraphics[width = .475\textwidth]{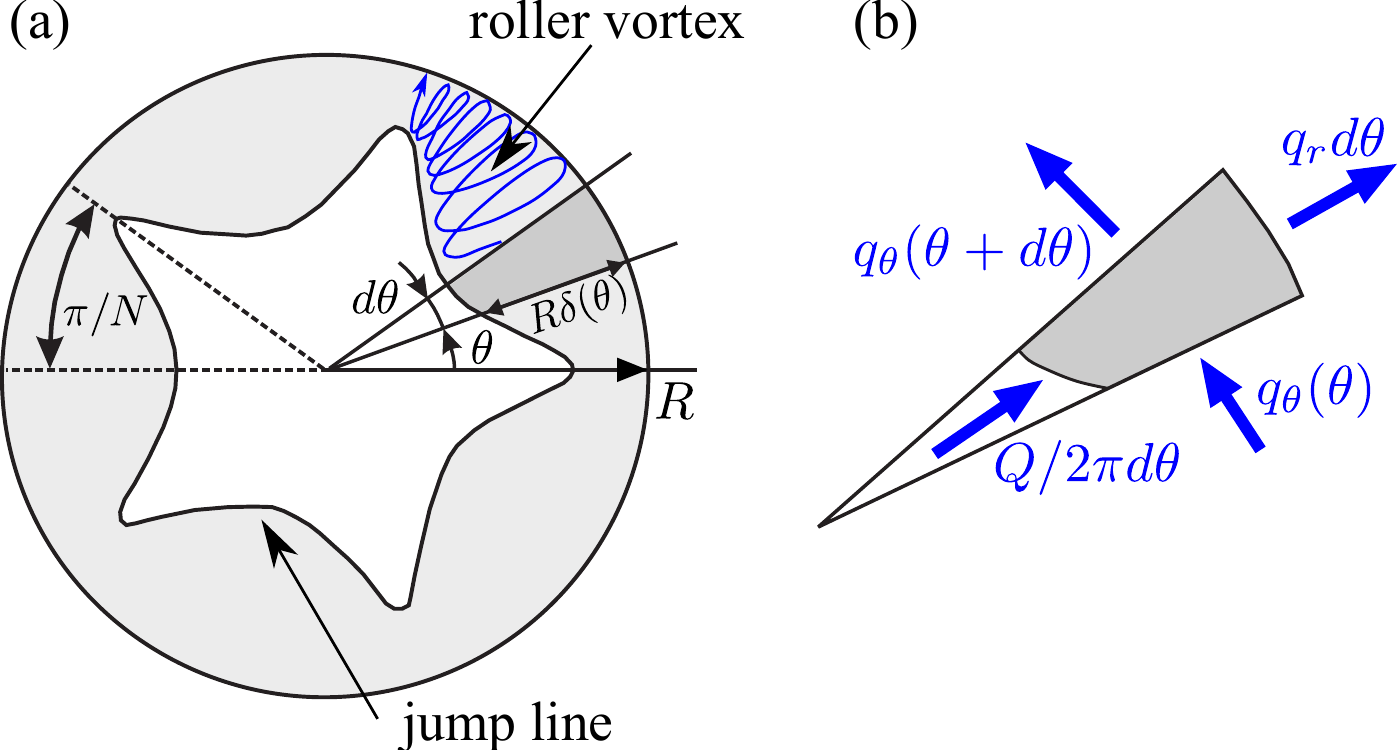}
    \caption{(Color online) 
(a) Schematic top view of a pentagonal jump. 
The gray shaded wedge represents the control volume with infinitesimal angle $d\theta$, also seen in Fig.~\ref{fig:jumpanatomy}.
(b) Enlargement of the wedge-shaped control volume. 
Arrows (blue) represent the radial and tangential mass flux $q_r$ and $q_{\theta}$, i.e. 
the flux in and out of the control volume.
\label{fig:massconserv}}
\end{figure}

FIG.~\ref{fig:massconserv} (a) shows the top view of the entire roller, and Fig.~\ref{fig:massconserv} (b) displays  an angular section $d\theta$ of the roller, together with the mass flux entering and leaving the control volume (corresponding to the gray shaded volume in FIG.~\ref{fig:jumpanatomy}).
Since the flow before the jump is purely radial and independent of $\theta$, as evidenced in FIG.~\ref{fig:heightprofile}, 
the radial flow into the volume must be $\frac{Q}{2\pi} d\theta$.
The fluid going out of the volume in the radial direction is denoted by $q_r(\theta) d\theta$.
The fluid going into the roller at the cross sectional wall at $\theta$ is $q_\theta(\theta)$, and the corresponding flow out of the roller at $\theta + d\theta$ is $q_\theta(\theta + d\theta)$.
Mass conservation in the region then demands
\[
    \frac{Q}{2\pi}d\theta + q_\theta(\theta) =
        q_r(\theta) d\theta + q_\theta(\theta + d\theta) .
\]
We define the normalized radial and azimuthal flux per azimuthal angle
$\xi_r \equiv q_r/Q$ and
$\xi_\theta \equiv q_\theta/Q$, respectively.
The continuity equation is now written as
\begin{eqnarray}
    \frac{1}{2\pi}& = &
        \xi_r(\theta) + \frac{d \xi_\theta}{d\theta}.\
\label{eq:masscons}
\end{eqnarray}
By integrating from $\theta=0$ to $2\pi$ and using the periodic condition for $\xi_\theta$,
we obtain a constraint for the total flux out of the jump region.
\begin{equation}
   \oint \xi_r(\theta) d\theta = 1 .
\label{eq:totalmasscons}
\end{equation}

\subsection{Radial Force Balance}
\label{sec:radialforcebalance}

\begin{figure}[ht]
\begin{center}
   \includegraphics[width=0.4\textwidth]{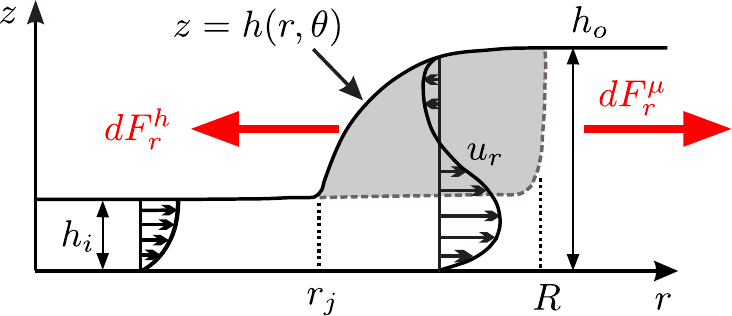}\\
   \caption[Velocity profile across the jump.]{(Color online) 
Side view of the wedge in Fig.~\ref{fig:jumpanatomy}.
The height inside and outside of the jump are $h_i$ and $h_o$, respectively.
We treat the roller eddy (shaded region) as a separate body of fluid
from the main stream, and assume the force balance of 
the hydrostatic force $dF_r^h$ and the shear force $dF_r^\mu$
in the radial direction (forces indicated by bold red arrows).
   \label{fig:vprofile}}
\end{center}
\end{figure}

We now inspect FIG.~\ref{fig:vprofile}, the side view of FIG.~\ref{fig:jumpanatomy}, and construct a force balance equation in the radial direction
for the roller which we view as a stationary body of liquid.
It experiences a force directed radially inward
due to the difference in fluid height at its inner and outer rim.
The shear force also acts on the bottom of the roller,
caused by the velocity gradient of the radially outward flow
along its bottom boundary which we estimate at $z=h_i$.
We assume that these forces balance, and that surface tension plays a secondary role in the radial direction.
Under this assumption, we have
\begin{equation}
    dF_r^h + dF_r^\mu = 0,
\label{radialforcebalance}
\end{equation}
where $dF_r^h$ and $dF_r^\mu$  are, respectively, the hydrostatic pressure force and the radial component of the viscous friction force per unit azimuthal angle on the roller slice in FIG.~\ref{fig:jumpanatomy}. 

\subsubsection{Hydrostatic pressure force.}
The pressure force, $dF_r^h$, is obtained by integrating the pressure over the area of its outer rim assumed to be at $r = R$ and extending from $z=h_i$ to $h_o$ (see Fig.~\ref{fig:vprofile}). Further, assuming the pressure to be hydrostatic, and omitting the ambient pressure which cancels, the force becomes
\begin{equation}
\label{hyd}
  dF_r^h = -\int_{h_i}^{h_o}\rho g (h_o-z)  \,dz\, R d\theta
  \approx -\rho g R \frac{h_o^2}{2} d\theta,
\label{radialhydrostaticforce}
\end{equation}
using (\ref{eqn:deltah}) where $\rho$ is the fluid density and $g$ is the gravitational acceleration.

\subsubsection{Shear.}
If the fluid is Newtonian with the dynamic viscosity $ \mu $, the outward force on the roller $dF_r^{\mu}$ shown in Fig.~\ref{fig:vprofile} is obtained by integrating the shear
$\tau_{rz} \approx \mu \left. \frac{\partial u_r}{\partial z} \right|_{z=h_i}$
over the bottom area of the roller:
\begin{equation}
  dF_r^\mu \approx - \int_{r_j}^R \mu 
  \left. \frac{\partial u_r}{\partial z} \right|_{z=h_i} rdr d\theta, 
\end{equation}
where $u_r$ is the radial velocity component.
To estimate this integral, we assume that the thin radial flow continues forward under the roller in a height of the order $h^*$,
an appropriate characteristic height near the roller.
The velocity in this region is related to the radial flux $q_r$
by $q_r = u_r^* r^* h^*$
where $r^*$ and $u_r^*$ denote some characteristic radius
and radial velocity, respectively.
We also assume that the shear can be expressed as
\[
   \left. \frac{\partial u_r}{\partial z} \right|_{z=h_i} \approx
   - c_1 \frac{u_r^*}{h^*} = - \frac{c_1 q_r}{r^* {h^*}^2}\, ,\
\]
with a numerical factor $c_1 >0$ of order unity.
Then, we have
\begin{equation}
  dF_r^\mu \approx
  \frac{c_1 \mu q_r}{r^* h^{*2}} d \theta \int_{r_j}^R r dr
  = \frac{c_1 \mu q_r}{2 r^* h^{*2}} (R^2 - r_j^2) d\theta .\
\end{equation}

For simplicity, we choose $r^* = \frac12 (r_j + R)$ and $h^*=h_i$.
Then the shear force becomes
\begin{equation}
  dF_r^\mu \approx \frac{c_1 \mu Q \xi_r}{h_i^2} (R - r_j) d\theta
  = \frac{c_1 \mu Q R}{h_i^2} \xi_r \delta d \theta .\
\label{radialshearforce}
\end{equation}

\subsubsection{Radial balance.}

Substituting the estimated forces
(\ref{radialhydrostaticforce}) and (\ref{radialshearforce}) into Eq.~(\ref{radialforcebalance}), we obtain the radial force balance equation
\begin{eqnarray}\label{eq:radly_balance}
    \xi_r = \frac{\Pi_1}{\delta},
\end{eqnarray}
with the nondimensional parameter
\begin{eqnarray}
     \Pi_1 \equiv \frac{g h_i^2 h_o^2}{2 c_1 \nu Q},
\end{eqnarray}
where $\nu$ is the kinematic viscosity $\mu / \rho$.
This equation was first derived  in \cite{Nature}. It shows that the radial flux increases at the corners where 
$\delta$ decreases. For a very thin corner, the flux would increase to infinity, and although our model remains meaningful in this limit, we do not expect it to represent the physics correctly, since the detailed structure of the corner flow could play an important role. Lacking such a more detailed model, we shall thus keep in mind that polygons with very thin or vanishing rollers in the corners are probably somewhat idealized.

In the circular state the normalized radial flux is $\xi_r = 1/2 \pi$ so the dimensionless roller has width $\delta_0$ where
\begin{equation}
\delta_0 = 2 \pi  \Pi_1.
\end{equation}

Combining (\ref{eq:radly_balance}) with the mass conservation (\ref{eq:masscons}), we obtain
\begin{equation}
\label{s2}
  {\frac{1}{2 \pi}} = {\frac{\Pi_1}{\delta}} + \xi'_{\theta}(\theta) ,
\end{equation}
whereas prime denotes the derivative with respect to $\theta$.
The constraint (\ref{eq:totalmasscons}), along with Eq.~(\ref{eq:radly_balance}), becomes
\begin{equation}
  \oint \frac{d\theta}{\delta (\theta)} = \frac{1}{\Pi_1} .
\label{eq:totalmasscons2}
\end{equation}

\subsection{Azimuthal Force Balance}
\label{sec:azimuthalmomentumbalance}
We now discuss the azimuthal transport in the control volume shown in FIG.~\ref{fig:massconserv}.
Since the flows in the azimuthal direction are slow, we neglect the kinetic term arising from the second term on the left hand side of (\ref{eq:momentumbalance}),
and consider hydrostatic pressure, viscous, and surface tension terms.

\subsubsection{Kinetic term.}
The second term on the left hand side of (\ref{eq:momentumbalance}) is the momentum flux through the boundary of the roller slice.
We consider the flux for the vertical sides of the cross sections at $\theta$ and $\theta + d\theta$ in FIG.~\ref{fig:jumpanatomy}.
Using the approximation $h_o\gg h_i$, the areas of these sides are given by
\begin{equation}
A_t(\theta) \approx  R \Delta h  \delta(\theta)  \approx R h_o  \delta(\theta) .
\label{At}
\end{equation}
We consider the average azimuthal flow velocity over such a surface
$\overline{u_\theta} = q_\theta/A_t$.
Then, the kinetic term is estimated as
\begin{eqnarray}\label{eq:dF_t_kin}
    dF_\theta^{kin} & = &
    \left[ \rho \overline{u_\theta}^2 A_t \right]_{\theta}^{\theta + d\theta}
    \approx     \rho \overline{u_\theta}^2 R h_o  \delta(\theta) d\theta\
\end{eqnarray}
Since the mean velocities in the roller always remain small, except, perhaps inside a very thin corner (where our model breaks down anyway, since the roller breaks) we shall neglect this term in the following.

\subsubsection{Hydrostatic pressure.}
For the azimuthal component of the first term on the right hand side of Eq.~(\ref{eq:momentumbalance}),
we assume the pressure in the roller to be hydrostatic.
Then, a net azimuthal pressure force on the roller slice arises
due to the difference in area of the two vertical sides at $\theta$ and $\theta + d\theta$ shown in Fig.~\ref{fig:jumpanatomy}.
The force on one side is estimated as in Eq.~(\ref{hyd}).
For this, we assume a simple surface profile which is linear within the cross-section $\theta=$ const., i.e.,
\begin{equation}
\label{h0}
  h(r, \theta) \approx h_i + {\frac{\Delta h}{R \delta(\theta)}}(r-R+R\delta(\theta))  
\end{equation}
where $h=h_i$ at $r=r_j(\theta)=R(1-\delta(\theta))$, and $h=h_o$ at $r=R$. With $h_o \gg h_i$ we get
\begin{equation}
\label{h}
  h(r, \theta) \approx  {\frac{\alpha}{\delta(\theta)}}[r-R(1-\delta(\theta))]  
\end{equation}
with the aspect ratio defined in (\ref{alpha}).
We then get
\begin{eqnarray}
  F_\theta^h (\theta) & = & \int_{R(1-\delta(\theta))}^R dr \int_{0}^{h(r,\theta)} \rho g \left(h(r,\theta)-z\right) dz 
\nonumber \\
  & \approx & \frac{\rho g R h_o^2}{6}\, \delta(\theta) .
\label{hyd1}
\end{eqnarray}
Considering the directions of the forces acting on the sides,
the force difference between the two sides is
\begin{equation}
\label{g}
  dF_\theta^h = - \frac{d}{d\theta} F_\theta^h (\theta) d \theta
  \approx - \frac{\rho g R h_o^2}{6} \,\delta'(\theta)d\theta,
\end{equation}
where prime again denotes the derivative with respect to $\theta$.
Due to the minus sign, the force acts in the direction from a belly to a corner of the roller.

\subsubsection{Shear.}
The contributions to the azimuthal component of the second term on the right hand side of Eq.~(\ref{eq:momentumbalance}) are the shear forces between the azimuthal flow inside the roller and the radial flow underneath and behind it. Just as for the radial force balance we must make some estimate of the velocity gradients involved.
The outer rim of the roller has area
\begin{equation}
  \label{eq:dAr}
  dA_r  \approx  R \Delta h d\theta \approx R h_o d\theta
\end{equation}
on which we estimate 
$\left. \partial u_\theta/\partial r \right|_{r=R}
\approx -c_2 \overline{u_\theta}/(R-r_j)$
since $u_\theta \approx 0$ for $r>R$.
Here, $c_2$ is a positive numerical factors of order unity, and 
$\overline{u_\theta}$ is the average azimuthal flow velocity over $A_t$.
The areas of the sides are given by
(\ref{At}).
Thus,
\begin{equation}
  \overline{u_\theta} = q_\theta/A_t \approx q_\theta/(R h_o \delta) .
\label{averageazimuthalvelocity}
\end{equation}

Likewise, the area on the bottom of the roller is
\begin{equation}
\label{eq:dAb}
  dA_b \approx \frac12(R^2-r_j^2)d\theta 
\end{equation}
on which $\left. \partial u_\theta/\partial z \right|_{z=h_i}
\approx c_3 \overline{u_\theta}/h^* \approx c_3 \overline{u_\theta}/h_i$
since $u_\theta = 0$ on $z=0$.
Here, we have used the height $h_i$ rather than $\Delta h$ to estimate the shear
between the roller and the main flow,
and $c_3$ is another positive numerical factor of order unity.

Note that the radial velocity gradient at the rim is measured from the interior to the exterior
of the roller, whereas the velocity gradient on the bottom is in the opposite direction.
Thus, we reverse the sign of the latter in order to add both contributions
for an estimate of the shear force on the control volume:
\begin{eqnarray}\label{eq:dF_t_mu}
    dF_\theta^\mu & \approx &
    - \mu \overline{u_\theta} \left[\frac{c_2}{R-r_j} R h_o
    + \frac{c_3}{h_i}\frac{R^2-r_j^2}{2} \right]d\theta \nonumber \\
    & = & - \frac{\mu q_\theta}{R h_o \delta} \left[ \frac{c_2 h_o}{\delta}
    + \frac{c_3 R \delta (R+r_j)}{2 h_i}\right]d\theta \nonumber \\
    & = & - \mu Q \xi_\theta \left[\frac{c_2}{R\delta^2}
    + \frac{c_3 R(2-\delta)}{2 h_i h_o}\right]d\theta.
\nonumber
\end{eqnarray}
Here, the relation (\ref{averageazimuthalvelocity}) is again used.

\subsubsection{Surface tension.}
\label{SurfaceTension}

Surface tension acts by changing the pressure inside the roller depending on the local curvature. This does not to a first approximation change the radial force balance, so we shall concentrate on the effects in the azimuthal direction. A somewhat similar approach in a different context can be found in \cite{Duclaux}. To model this, we shall include the Laplace pressure difference across the surface $z=h(r,\theta)$:
\begin{equation}
\label{e1}
\Delta p = \sigma \left({\frac{1}{R_1}} + {\frac{1}{R_2}} \right),
\end{equation}
where $R_1$ and $R_2$ are the two principal curvatures;
we take their signs to be positive if the center of curvature is on the ``inside'' direction from the surface (i.e.\ located under the surface).
Taking the atmospheric pressure to be zero everywhere,
the relation (\ref{e1}) describes the fluid pressure just beneath the surface.
Thus, the total pressure inside the fluid becomes
\begin{equation}
  p (r,\theta,z) = \rho g (h(r, \theta) -z) + \Delta p .
\label{totalpressure}
\end{equation}
To use these expressions we need information about $h(r,\theta)$, in the circular state and in the polygonal state --- and in the family of states between them. Of course, we do not have this information, so, in keeping with our simple model, which expresses the geometry of the jump entirely in terms of the local width of the roller, we shall replace the two radii of curvature with averages over the cross-section of the surface.

In what follows, we consider small perturbations, i.e. nearly circular jumps.
One principal radius of curvature $R_1$ quantifies the curvature roughly in the $(r,z)$-plane, i.e., the cross section of the jump. This is the ``dangerous" one, which may lead to instability. The other one, $R_2$, is defined in a plane orthogonal to $R_1$, also including the surface normal.
Since surface inclinations are small, $R_2$ can be approximated by the radius of curvature of the jump shape seen from above (cf. Fig.~\ref{fig:massconserv}).
For slow variations of the plane curve $R \delta(\theta)$ the radius of curvature we get
\begin{equation}
\label{R2} 
 {\frac{1}{R_2}} \approx -  {\frac{1}{R}} \delta''(\theta),
\end{equation}
where primes denote differentiation with respect to $\theta$.

To estimate $R_1$  we have to know how the shape distorts when a corner is created. Since it is observed that addition of surfactant destroys the polygons (see also Supplemental Material~\cite{suppmaterial}), it seems likely that surface tension can make the circular states unstable through a Rayleigh-Plateau-like instability (however, it cannot explain the observation of circular type-II states that are sometimes stable.).
This means that the curvature should increase when $\delta$ is decreased as it does near a corner. This is indeed observed in experiments as seen in Fig.~\ref{fig:heightprofile}, where the height profile (on a radial path) of a polygon-state has been measured with a conducting needle~\cite{1999ma}, both through an edge and a corner. It is clear that the curvature is larger (i.e.\ the radius of curvature is smaller) through the corner.
Thus, we assume 
\begin{equation}
\label{R111}
R_1'(\delta) >0
\end{equation}
where prime denotes differentiation with respect to $\delta$.


When the height $\Delta h$ and the width $R \delta$ are similar, we expect that $R_1$ would have a value close to these, like for a cylindrical surface, but in contrast to the classical cylinder case of Plateau and Rayleigh, the curvature cannot keep increasing when the roller shrinks since the external height is fixed.
Thus, we expect the curvature to saturate when the width of the roller becomes much smaller than $\Delta h$. For future use, we shall use the symbol $\rho_1$ for the normalized cross-sectional curvature: $\rho_1 = R_1/R$.
The Laplace pressure term in (\ref{totalpressure}) now becomes
\begin{equation}
  \label{deltap}
  \Delta p  = \frac{\sigma}{R} \left( {\frac{1}{\rho_1(\delta)}} - \delta'' \right).\
\end{equation}

To get the force, we now have to integrate $\Delta p \cdot {\mathbf{\hat n}}$ over the bounding
surfaces, where ${\mathbf{\hat n}}$ is the local outward normal, and project the result along $\theta$. This will give the force from
the roller on the surroundings. To get the force {\em on} the roller we have
to put in a minus-sign. The rear surface does not contribute
anything since it is parallel to ${\hat \Theta}$. Treating $\Delta p$
as constant in the cross-section $\theta =$ const. yields
\begin{equation}
  F_{\text{\tiny vert}, \sigma}\approx -A_t \Delta p \,{\hat \Theta}
\label{forcesigmavertical}
\end{equation}
where the areas of the sides $A_t$ are given by (\ref{At}).

To estimate the free surface contribution we need the area of the free surface and the $\theta$-component of the outward unit normal vector. As above, we use the linear height profile (\ref{h}), where
\begin{equation}
\label{h'1}
h,_{r} \equiv \frac{\partial h}{\partial r} \approx \frac{\Delta h}{R \delta}
\end{equation}
and
\begin{equation}
\label{h'2}
h,_{\theta} \equiv \frac{\partial h}{\partial \theta} \approx {\frac{\delta'(\theta) \Delta h}{ \delta(\theta)^2}} \left(1-\frac{r}{R} \right) .
\end{equation}
where, for brevity, we use a compact notation for partial derivatives throughout this section.
The area of the free surface is
\begin{equation}
  dA_f \approx R d \theta \sqrt{(R \delta)^2 + (\Delta h)^2},
\end{equation}
and the unit normal vector is 
\begin{equation}
{\mathbf{\hat n}}= N \left(
\begin{array}{c}
-h,_{r} \\ 
-h,_{\theta}/r\\
1\
\end{array}
\right)
\end{equation}
with a suitable normalization factor $N$.
Assuming that $\delta'(\theta) \ll 1$ so that $| h,_{\theta}/r | \ll h,_{r}$ and $| h,_{\theta}/r | \ll 1$, we have
\begin{equation}
N \approx \frac{1}{\sqrt{(h,_{r} )^2+1}} = {\frac{R \delta}{\sqrt {(\Delta h) ^2+(R \delta )^2}}} .
\end{equation}
The $\theta$-component of the unit normal becomes
\begin{equation}
{\hat n}_{\theta}  \approx -{\frac{N h,_{\theta}}{R }}   \approx -  {\frac{\delta \,h,_{\theta}}{\sqrt {(\Delta h )^2+(R \delta )^2)}}}
\end{equation}
and we obtain (based on the assumption $\Delta h \ll R \delta$)
\begin{equation}
dA_f \, {\hat n}_{\theta} \approx  -R\, \delta\, h,_{\theta}\; d \theta .
\end{equation}
Using the average value for $h,_{\theta}$,
\begin{equation}
\overline{h,_{\theta}} = {\frac{1}{R \,\delta(\theta)}} \int_{R(1-\delta(\theta))}^R h,_{\theta} \;dr = 
{\frac{1}{2}} {\frac{\Delta h}{ \delta(\theta)}}  \,\delta'(\theta),
\end{equation}
we obtain
\begin{equation}
  dA_f \, {\hat n}_{\theta} \approx - {\frac{1}{2}} R \,h_o \,\delta' d \theta .
\end{equation} 
The force contribution from the free surface is this product times $\Delta p$.
Thus,
\begin{eqnarray}
\nonumber
  d F_{\text{\tiny free}, \sigma} &\approx& -dA_f {\hat n}_{\theta} \Delta p \\
  & \approx & {\frac{\sigma h_o}{2}} \delta' \left( {\frac{1}{\rho_1(\delta)}} -\delta'' \right) d \theta .
\label{forcesigmafree}
\end{eqnarray}
The total surface tension force in the azimuthal direction can now be computed from (\ref{forcesigmavertical}) and (\ref{forcesigmafree}).
Note that we need the forces {\it on} the roller, so the direction of the force is {\it into} the roller, i.e., in the direction ${\hat \theta}$ at $\theta$ and in the direction $-{\hat \theta}$ at $\theta + d\theta$. Thus,
\begin{eqnarray}
  d F^\sigma_\theta & \approx & - F'_{\text{\tiny vert},\sigma}(\theta) d \theta + d F_{\text{\tiny free}, \sigma} 
\nonumber \\
 & \approx &
 {\frac{1}{2}} h_o  \, \sigma \delta \left[ {\frac{ \delta' \rho_1'(\delta)}{\rho_1^2}} + \delta'''(\theta) \right] .
\end{eqnarray}
It is seen that if $\rho_1'(\delta_0) > 0$ (like in the case of a cylinder squeezed uniformly), the first term gives a contribution similar to the hydrostatic one (\ref{g}), but  with {\it opposite} sign. This seems to be the case for the polygonal hydraulic jump shown in FIG.~\ref{fig:heightprofile}, where the curvature in the corner is larger than at the edge.

For later use, we shall introduce the shape parameter
\begin{equation}
\label{B}
 B \equiv {\frac{\delta_0 \rho_1'(\delta_0)}{(\rho_1(\delta_0))^2}},
\end{equation}
For a cylindrical roller with $\rho(\delta) \approx \delta $ we would then get $B \approx \delta_0^{-1}$.
The ratio of the prefactor of the gravitational term (\ref{g}) to the surface tension term (without $\delta$) is the dimensionless number
\begin{equation}
\label{Bond0}
{\frac{\rho g\, R \,h_o^2/2}{3 \sigma \,h_o/2}} =  {\frac{R h_o}{3 l_c^2}} =\frac{\rm Bo^2}{3 \alpha}
\end{equation}
where $l_c = [\sigma/(\rho g)]^{1/2}$ is the capillary length and where
\begin{equation}
\label{Bond}
{\rm Bo} = \frac{h_o}{l_c}
\end{equation}
is the Bond number based on $h_o$. 

\subsubsection{Azimuthal force balance.}
\label{sec:azimuthalforcebalance}
Using these estimates, we write down
the azimuthal component of Eq.~(\ref{eq:momentumbalance}):
\begin{equation}
  dF_\theta^h + dF_\theta^\mu + dF_\theta^\sigma = 0 ,
\label{azimuthalforcebalance}
\end{equation}
which can be written in  dimensionless form as
\begin{eqnarray}
\nonumber\label{eq:staticEq1}
\delta'(\theta)
    & = & -\left[
      \frac{\Pi_2}{\delta^2} + \Pi_3\left (1-\frac{\delta}{2}\right) \right] \xi_\theta \\
      \label{s1}
    &+ & 3 \alpha {\rm Bo^{-2}} \delta \left({\frac{\rho_1'(\delta)}{\rho_1^2}} \delta'(\theta)+    \delta'''(\theta)  \right),
\end{eqnarray}
where we have defined the dimensionless numbers
\begin{equation}
\label{PI}
    \Pi_2 \equiv \frac{2 c_2 \nu Q}{g R^2 h_o^2} , \quad \mbox{and} \quad
    \Pi_3 \equiv \frac{2 c_3 \nu Q}{g h_i h_o^3} .
\end{equation}

The two coupled Equations (\ref{eq:staticEq1}) and (\ref{s2}) close our model for $\delta$ and $\xi_\theta$.
An overview of the numerous symbols used in our model is provided in Table~\ref{nomenclature} in Sec.~\ref{appendixB}.

\subsection{Linear analysis}
\label{sec:linearization}
We investigate the existence of nearly circular polygons, i.e. for small oscillations around the circular solution $\delta = \delta_0= 2 \pi \Pi_1 $ and $\xi_{\theta} =0$.
Inserting $\delta = \delta_0 + \epsilon \delta_1(\theta)$ and $\xi_\theta = \epsilon \xi_1(\theta)$ into (\ref{s1}) and (\ref{s2}), and by using $B$ in (\ref{B}), we find the first order perturbation equations to be

\begin{eqnarray}
\nonumber\label{eq:LinAnalysis1a}
(1- 3 \alpha {\rm Bo^{-2}} B ) \;\delta_1 ' & = & - \xi_1\left[\frac{\Pi_2}{\delta_0^2}+\Pi_3(1-\frac12\delta_0)\right] \\
    & +& 3 \alpha {\rm Bo^{-2}}  \delta_0  \;\delta_1 ''' ,\\\label{eq:LinAnalysis1b}
    \xi_1' & = &{ \frac{1}{2 \pi \delta_0}}\delta_1.
\end{eqnarray}
Differentiating (\ref{eq:LinAnalysis1a}) with respect to $\theta$ and substituting $\xi_1'$ from (\ref{eq:LinAnalysis1b}), we obtain
\begin{eqnarray}
\nonumber
      \left( 1 - 3 \alpha {\rm Bo^{-2}} B   \right) \delta_1 '' & = &  \frac{1}{2 \pi \delta_0}\left[\frac{\Pi_2}{\delta_0^2}+\Pi_3(1-\frac12\delta_0)\right]\delta_1\\
      &+&  \delta_0    3 \alpha {\rm Bo^{-2}}  \; \delta_1''''.
\end{eqnarray}
For perturbations with $\delta_1=\sin(k(\theta-\theta_0))$,
we obtain the characteristic equation:
\begin{eqnarray}
\nonumber
\delta_0 k^4 &+& k^2 \left( \frac{{\rm Bo^{2}}}{3 \alpha}- B   \right)  \\
& - &  \frac{{\rm Bo^{2}}}{6 \pi \alpha \delta_0}\left[\frac{\Pi_2}{\delta_0^2}+\Pi_3\left(1-\frac12\delta_0\right)\right]=0.
\end{eqnarray}
This equation has the solution 
\begin{equation}\label{eq:klin}
 k^2 =\frac{1}{2 \delta_0} \left(B- \frac{{\rm Bo^{2}}}{3 \alpha} \pm \sqrt{G }     \right),
\end{equation}
where we define
\begin{equation}
G\equiv\left( B -  \frac{{\rm Bo^{2}}}{3 \alpha}\right)^2 + \frac{2 {\rm Bo^{2}}}{3 \alpha \pi}\left[\frac{\Pi_2}{\delta_0^2}+  \Pi_3\left(1-\frac12\delta_0\right)\right].
\end{equation}
When the wave number $k$ is an integer, the solution is periodic on $[0,2 \pi/ k]$ and corresponds to a polygon with $k$ corners. 
In the present case, $k^2$ needs to be positive for the solution to make sense. In the limit of negligible surface tension where $\sigma\rightarrow 0$ (i.e. $ {\rm \Bo \rightarrow \infty} $),
we simply get
\begin{equation}
\label{kk}
 k= \pm \sqrt{  \frac{1}{2 \pi \delta_0}\left[\frac{\Pi_2}{\delta_0^2}+\Pi_3\left(1-\frac12\delta_0\right)\right] }.     
\end{equation}
\section{Solvable Nonlinear Model}
\label{sec:case1}



To solve the static nonlinear model, Eqs.~(\ref{s1})-(\ref{s2}), we would need to know $\rho_1(\delta)$. At present we do not know the shapes well enough to estimate this, and we shall restrict our attention to the case where surface tension is absent; despite this simplification, we shall see that we still obtain meaningful results which allow us to explain the structure of the phase diagram.

It is instructive to start with a further simplification. By letting $\Pi_3\rightarrow 0$, we neglect the term due to the shear against the bottom of the roller.
By ignoring this regular perturbation term, the model can be solved analytically, which enables further analysis, as discussed in Sec. \ref{sec:Pi3Model}, where we investigate the effect of the $\Pi_3$ term. We obtain
\begin{eqnarray}
\label{eq:goveqncase1}
   \phi^2 x' &=& -\frac{y}{x^2},\\
   \label{eq:goveqncase2}
   y' &=& 1- {\frac{1}{x}},
\end{eqnarray}
where we have introduced the rescaled variables
\begin{eqnarray}\label{eq:scaling1}
  x(\theta) &=& \frac{\delta}{\delta_0} \\
  y(\theta) &=& 2 \pi \xi_{\theta},\
\end{eqnarray}
and the non-dimensional parameter 
\begin{equation}
  \phi= (2 \pi)^2 \Pi_1^{3/2} \Pi_2^{-1/2} = \frac{\pi^2 h_i^3 h_o^{4} g^2 R}{\nu^2 Q^2 \sqrt{c_1^3 c_2}}.
\end{equation}
We now solve (\ref{eq:goveqncase1}) for $y$, differentiate it once with respect to $\theta$,
and eliminate $y'$
in (\ref{eq:goveqncase2}), i.e., to obtain the single (second-order) equation
\begin{equation}\label{eq:xgoveqn}
    \phi^2 \left( \frac{x^3}{3} \right)'' = \frac{1}{x} - 1 .
\end{equation}
By substituting $X = x^3/3$, this may be formulated as
an equation of motion for a ``mass'' $\phi^2$ in the conservative force field,
\begin{equation}
    \phi^2 X'' = - \frac{dV}{dX},
\end{equation}
where the potential function is given by
\begin{equation}\label{eq:potential}
  V = X - \frac12 (3X)^{2/3} = \frac{x^3}{3} - \frac{x^2}{2} .
\end{equation}
We introduce the rescaled angle $\Theta=\theta/\phi$ and integrate once in $\Theta$ to obtain
\begin{equation}
  \frac{1}{2} \left( \frac{dX}{d\Theta} \right)^2 + V(X) = C,
\end{equation}
where $C$ is a constant of integration corresponding to the ``energy'' of the mass $\phi^2$. Solving the equation for $dX/d\Theta$ and transforming back to $x$, we obtain
\begin{equation}
\label{eq:xprime}
  \frac{dx}{d\Theta} = \pm \frac{\sqrt{2 (C - V(x) )}}{x^2}.
\end{equation}
This equation has the implicit solution
\begin{equation}
    \Theta(x_b) = \Theta(x_a)  \pm \int^{x_b}_{x_a} \frac{x^2\,dx  }{\sqrt{2\left(C-V(x)\right)}}.
\end{equation}
With (\ref{eq:scaling1}) we recover the expression in terms of the original scaling
\begin{equation}
    \theta(\delta) = \theta(\delta_a) \pm \phi \int_{\delta_a/\delta_0}^{\delta/\delta_0} \frac{x^2\,dx  }{\sqrt{2\left(C -V(x)\right)}},
\end{equation}
and similarly, by substitution of (\ref{eq:xprime}) into (\ref{eq:goveqncase1}), the corresponding azimuthal flux is given by
\begin{equation}
 \xi_\theta = \mp \frac{\phi}{2\pi } \sqrt{2(C-V(\delta/\delta_0))} .
\end{equation}

\subsection{Phase Diagram}
\label{sec:case1phasediagram}

\begin{figure}[tbh]
    \includegraphics[width=0.4\textwidth]{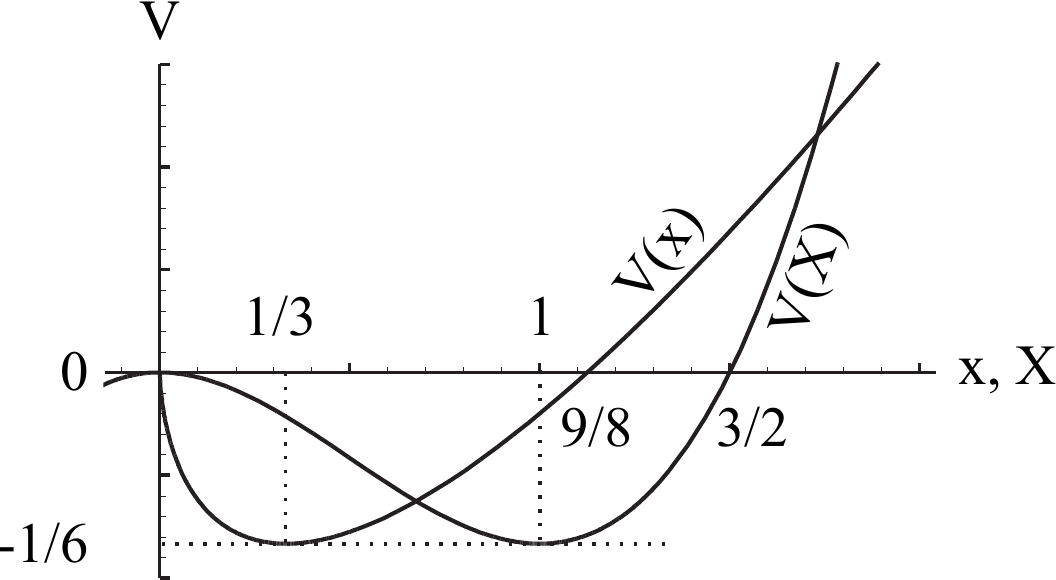}
\caption{
The potential function $V$ as a function of $x=\delta/\delta_0$ and $X=x^3/3$, respectively.
A periodic solution exists only when $-\frac16 < C < 0$.
The solution corresponds to a polygon with $N$ corners
if the period is $2\pi/N$ in terms of $\theta$.
\label{fig:phdiagcubic}}
\end{figure}

We now consider the parameter space for existence of polygons. Such solutions may be considered ``bound" states of the potential $V$. As shown in FIG.~\ref{fig:phdiagcubic}, such a state exists if and only if $-\frac16 < C < 0$. When $C$ is within the range, there are two roots $\xmin$ and $\xmax$ for the equation $V(x)=C$, satisfying $0 < \xmin < \xmax < \frac32$.
They serve as turning points for a trajectory, and describe the minimum (corner) and the maximum (belly) of the jump width, respectively.
Without loss of generality we may choose $x=\xmin$ when $\theta=\Theta=0$, and
then a trajectory can be computed by integrating (\ref{eq:xprime})
with the plus sign from $\Theta=0$ to $\Theta=T/2$,
where $T(C)$ denotes the period of oscillation in terms of $\Theta$,
and extending that part of the solution using symmetry.

For a polygon with $N$ corners a trajectory must oscillate
in the potential $N$ times,
and this must result in an increase by $2\pi$ in terms of $\theta=\phi \Theta$.
Thus, a solution for $N$-polygon must satisfy
a commensurability condition
\begin{equation}
  \phi N = \frac{2 \pi}{T(C)},
\label{eq:polygoncondition}
\end{equation}
with the normalized half period 
\begin{equation}
\label{eq:halfperiod}
   \frac{T(C)}{2} = \int_{\xmin(C)}^{\xmax(C)}
   \frac{x^2 dx}{\sqrt{2 ( C - V(x) )}} .
\end{equation}

When $C = -\frac16$ the equilibrium solution 
$\xmin = \xmax = 1$,
corresponding to the circular jump $\delta \equiv \delta_0$,
is obtained.
When $C $ is only slightly larger than $-\frac16$
we approximate the potential near the local minimum as
$V(x) \sim -\frac16 + \frac12 (x-1)^2$.
If we express $\xmin=1-\epsilon$, then $\xmax=1+\epsilon$
and the ``energy'' is approximated by $C = V(\xmin) = V(\xmax) \sim -\frac16 + \frac12 \epsilon^2$.
Then, $2 (C-V(x) ) \sim \epsilon^2 - (x-1)^2$ and we find
\begin{equation}
  \frac{T (-1/6)}{2} = \lim_{\epsilon\rightarrow 0} \int_{1-\epsilon}^{1+\epsilon}
  \frac{x^2 dx}{\sqrt{\epsilon^2 - (x-1)^2}} = \pi ,
\end{equation}
for which the condition (\ref{eq:polygoncondition}) becomes
$\phi = 1/N$.
Naturally, this agrees with the condition
(\ref{kk}) found in the linear analysis when $\Pi_3=0$.

From $C=-\frac16$ to $C = 0$,
the oscillation amplitude grows until, at $C = 0$,
$\xmin = 0$ and $\xmax = \frac32$.
Correspondingly, the integral $T(C)/2$
decreases monotonically in $C$ until it reaches
\begin{equation}
  \frac{T(0)}{2} = \int_{0}^{3/2} \frac{x^2 dx}{\sqrt{-2 V(x)}} = 3 .
\end{equation}
In this limit the condition (\ref{eq:polygoncondition}) becomes
$\phi = \pi/(3N)$.

To summarize, the condition (\ref{eq:polygoncondition})
for the existence of a polygon solution becomes
\begin{equation}
   \phi = \frac{K}{N},
\label{eq:case1bif}
\end{equation}
where $K = 2 \pi / T \in [1,\pi/3] \approx [1,1.0472]$.
$K=1$ corresponds to nearly circular and $K=\pi/3$ to spiky polygons.

\begin{figure}[tbh]
        \includegraphics[width=0.4\textwidth]{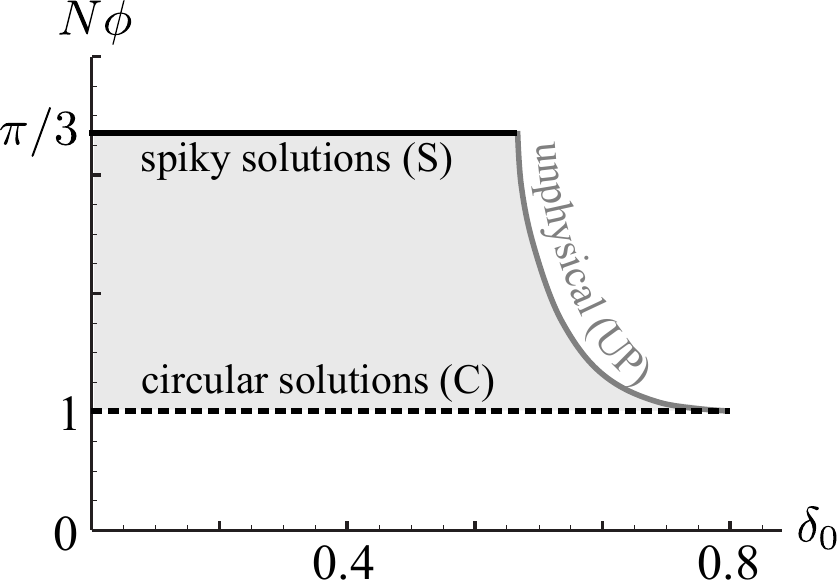}
\caption{
Polygonal jumps with $N$ corners exist in the parameter range shown above (truncated nonlinear model).
in the truncated nonlinear model.
The jumps bifurcate from the line $\phi=1/N$ for the circular jumps (C).
When the parameters approach the line $\phi=\pi/(3N)$, 
the roller thickness approaches $\delta=0$ in the corners at some $\theta$,
indicating a rupture;
solutions appear spiky near the line (S).
On the border $\delta_0 = 1/\xmax$,
the roller thickness reaches $\delta=1$ so the solution becomes unphysical (UP).
\label{fig:bd1a}}
\end{figure}

We also need to ensure $0< \delta <1$.
Since $\xmin>0$, $\delta>0$ is guaranteed.
Thus, the remaining condition is $\delta <1$ for all $\theta$
which is equivalent to
\begin{equation}
  \delta_0 < 1/\xmax .
\label{eq:case1bif2}
\end{equation}
On this border the roller thickness is predicted to equal the jump radius.
This unphysical behavior is likely to be due to truncation of the terms
in this approximation.
The border curve is computed in FIG.~\ref{fig:bd1a}
together with the lines $\phi = 1/N$ and $\pi/(3N)$.

Since $\xmax$ is determined purely by the choice of $C$,
the border (\ref{eq:case1bif2}) is identical for all $N$
but only scaled and shifted in the $\phi$-direction.
Apart from this border,
an $N$-polygon exists in the horizontal strip $\pi/(3N) > \phi > 1/N$.
It is natural to ask for which $N$ we find an overlap of these bands corresponding to multistability of polygons as observed experimentally. Consider two neighboring strips with $N$ and $N+1$.
The condition for overlap of the strips is $\phi_N^\textrm{\tiny min}<\phi_{N+1}^\textrm{\tiny max}$,
or $1/N<\pi/\{ 3(N+1) \}$.
This leads to a condition $N>3/(\pi-3)\approx21.2$, or $N\geq22$.
While the model thus in principle predicts overlapping regions
and resulting hystereses of polygonal jumps with different $N$,
it is an experimental fact that overlap exists for much lower $N$.
\subsection{Solutions and Shapes}
\label{sec:case1solutionsandshapes}

\begin{figure}[tbh] 
 \includegraphics[width=0.45\textwidth]{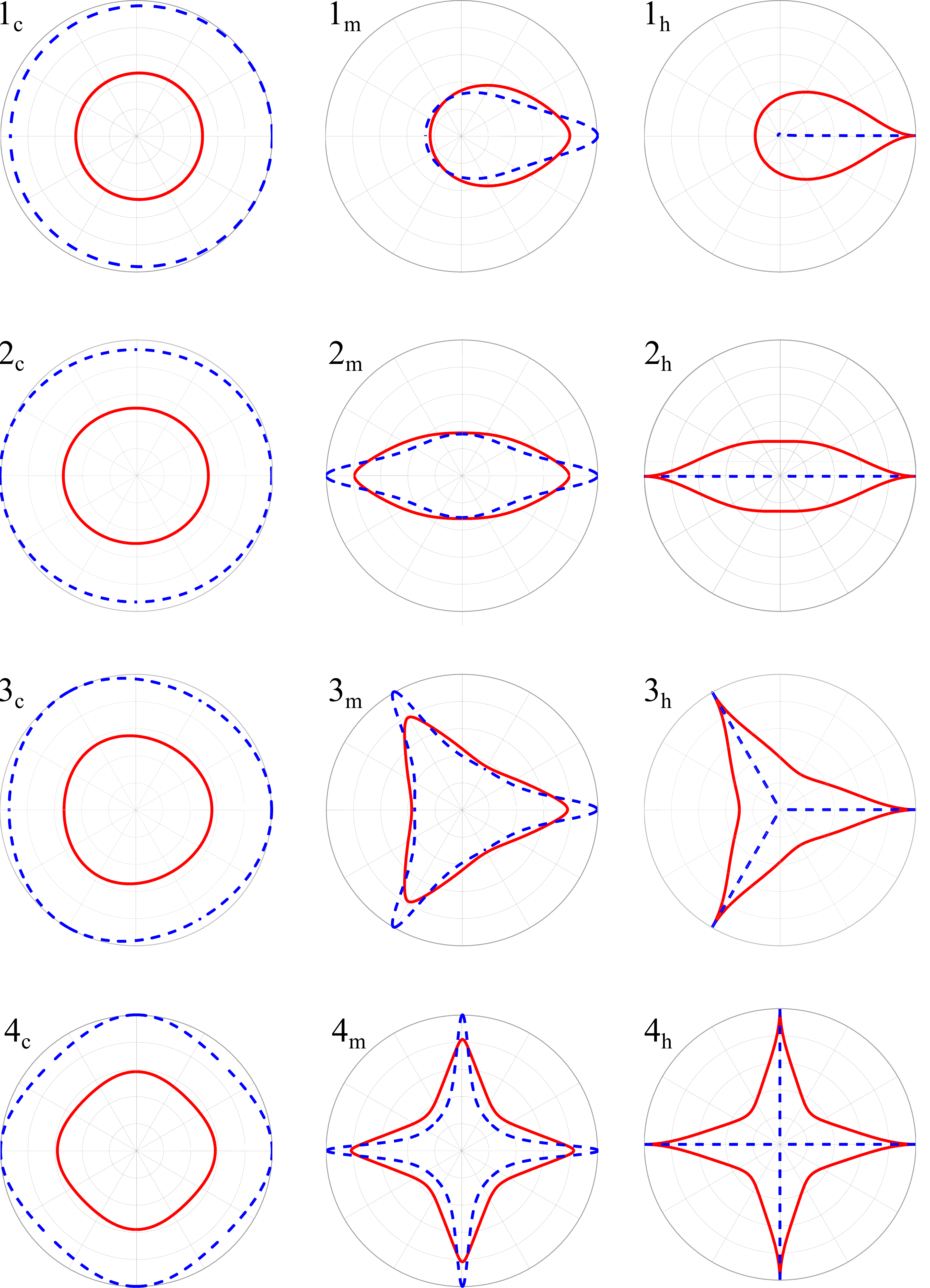}
\caption{(Color online) 
Typical polygon solutions are shown for the truncated model in 
Eqs.~(\ref{eq:goveqncase1})-(\ref{eq:goveqncase2}). The polygon shape (jump line) is given by 
$\delta(\theta)$ (solid red) and the radial flux by $\xi_r(\theta)$ 
(blue dashed), which is computed via the radial balance 
Eq.~(\ref{eq:radly_balance}). Numbers refer to the symmetry $N$ and 
increase down the columns. From left to right, the transition from 
nearly circular to spiky is shown for each symmetry ('c', 'm' and 'h' 
refer to center, middle or homoclinic orbit, respectively), 
corresponding to the transition discussed in  Fig.~\ref{fig:bd1a}.
\label{fig:1234shapes}}
\end{figure}

Some typical solutions are displayed in FIG.~\ref{fig:1234shapes} in terms of the variables $\delta(\theta)$ and $q_r(\theta)$ for the symmetries $N=1,2,3,4$, which were obtained by solving Eqs.~(\ref{eq:goveqncase1})-(\ref{eq:goveqncase2}) numerically. In Fig.~\ref{fig:1234shapes}, each row shows the transition from nearly circular to spiky solutions -- this may be achieved by tuning $N\phi$ between 1 and $\pi/3$. 
Notice that a solution only exists for a special value of the energy $C$, and is determined via Eq.~(\ref{eq:polygoncondition}) by the choice of the symmetry $N$ and ``mass'' $\phi^2$. The strict monotonicity of $T=T(C)$ in $C$ (see above) means that this mapping is one-to-one. It is meaningful to start integration in a corner where $x'=0$ at $\xmin$ or $\xmax$, respectively.
The initial condition $\xmin$ is then given by this $C$, (see Eq.~(\ref{eq:xprime}) which yields $V(x_\textrm{\tiny min,max})=C$ for $x'=0$). In other words, to obtain the desired symmetry $N$ with mass $\phi^2$, the initial condition $\xmin$ must be chosen accordingly.


The spiky polygons go along with a singular behavior of the radial outward flux $q_r(\theta)$ (blue).
This behavior is caused by $\delta \rightarrow 0$ in Eq.~(\ref{eq:radly_balance}),
and it is likely that this behavior disappears by inclusion of some of the neglected terms. However, we note that although this behavior is not physical, it is seen that the polygons exhibit very strong jets in the corners in the outward radial direction, as is seen in the flow visualization in Fig.~\ref{fig:rollerflowvisual}.
Notice also that the wave number has the right dependence on $h_o$, i.e. larger $h_o$ leads to fewer corners. Thus the increase in the number of corners with increasing $Q$  --- as seen in FIG.~\ref{fig:phasediagram} --- is reproduced.

\subsection{Comparison with experimental observations}
\label{sec:case1comparison}

We now represent the shaded region in FIG.~\ref{fig:bd1a} in terms of
the physical entities.
Our aim is to compare the predicted region with 
the measurement on the $(h_o,Q)$-plane in FIG.~\ref{fig:phasediagram}.
The two parameters in the model can be written as
\begin{eqnarray}
  \delta_0 & = & \pi g h_i^2 h_o^2 / (c_1 \nu Q)
\label{eq:delta0relation} \\
  \phi & = & \pi^2 g^2 R h_i^3 h_o^4 / (c_1^{3/2} c_2^{1/2} \nu^2 Q^2) .
\label{eq:phirelation}
\end{eqnarray}
We notice that both $\delta_0$ and $\phi$ appear to depend on
$h_o$ and $Q$ only through the combination $h_o^2/Q$.
If so, then we are not able to map the bifurcation curves from FIG.~\ref{fig:bd1a} one-to-one to the phase diagram curves in FIG.~\ref{fig:phasediagram}.

However, the jump radius $R$ actually depends on other parameters
as described before.
For an estimate of this dependence, we use the relation $R  = c_4 Q^{5/8} \nu^{-3/8} g^{-1/8}$
($c_4 = $ const.)\ proposed in \cite{Dimon},
since it is based on experimental data and leads to a simple estimate for the mapping.
This converts Eq.~(\ref{eq:phirelation}) to
$\phi \sim \pi^2 c_4 g^{15/8} h_i^3 h_o^4 / (c_1^{3/2} c_2^{1/2} \nu^{19/8} Q^{11/8})$,
and the $Q$-dependence of $\phi$ is corrected.
Solving this with (\ref{eq:delta0relation}), we obtain
\begin{eqnarray}
  h_o & \sim & \left( \frac{c_1 c_2^4 \nu^8 \phi^8}{\pi^5 c_4^8 g^4 h_i^2 \delta_0^{11}} \right)^{1/10}
\label{eq:horelation} \\
  Q & \sim & \left( \frac{c_2^4 g \nu^3 h_i^8 \phi^8}{c_1^4 c_4^8 \delta_0^{16}} \right)^{1/5} .
\label{eq:Qrelation}
\end{eqnarray}
Thus we are able to map $(\delta_0,\phi)$ back to the physical parameters $h_o$ and $Q$.

\begin{figure}[tbh]
    \includegraphics[width=0.475\textwidth]{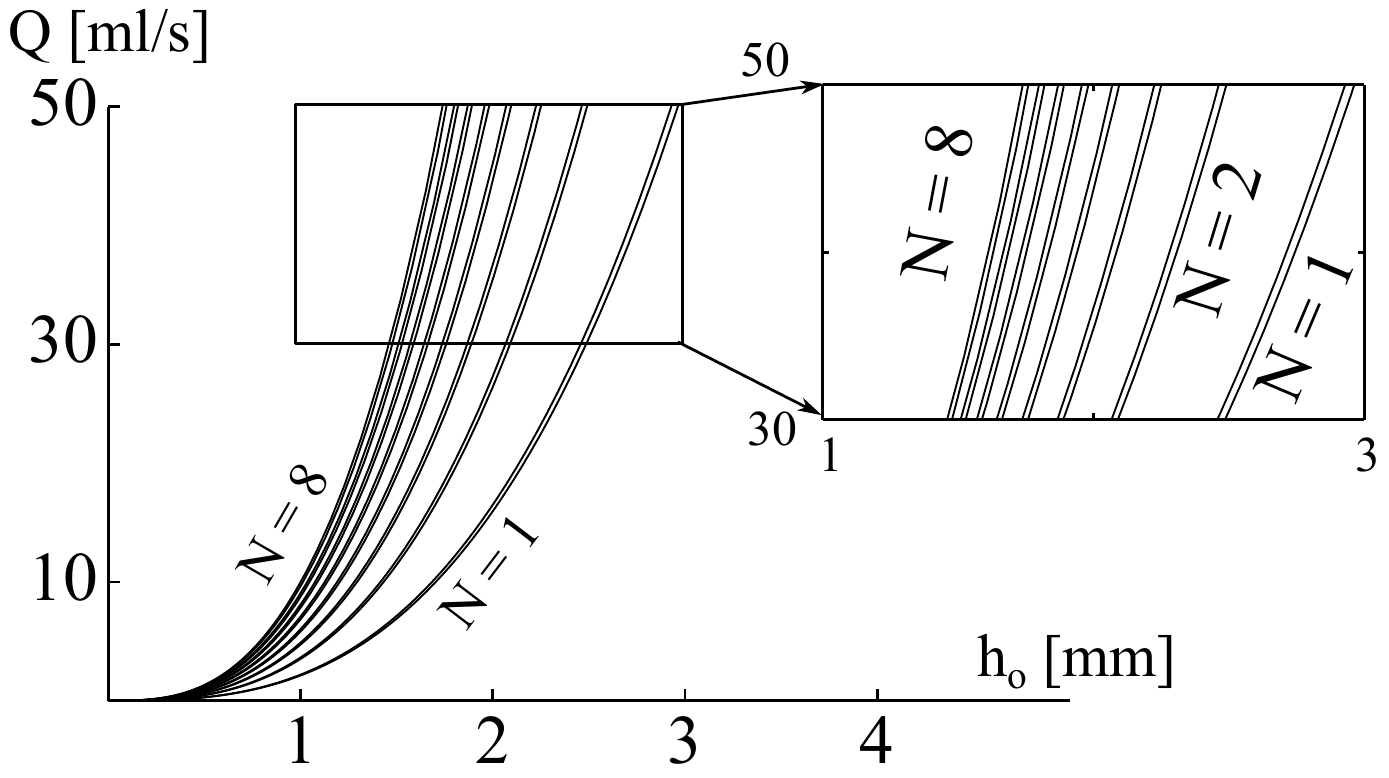}
    \caption{
Bifurcation diagram in physical parameters ($h_o,Q$). The regions in which polygonal solutions with each $N$ up to $N=8$ are shown.  The $h_o$-$Q$ relationship for the border curves, especially in the enlarged inset, is qualitatively similar to  the measured FIG.~\ref{fig:phasediagram} despite quantitative disagreement. The bands are predicted to be thinner and do not overlap in the model, unless $N \geq 22$, unlike in experimental observations.
\label{fig:physicalborder}}
\end{figure}

When only $\delta_0$ is varied while $\phi$ and other parameters
on the right hand sides of (\ref{eq:horelation},\ref{eq:Qrelation}) are fixed,
these relations predict
$h_o \propto \delta_0^{-11/10}$ and $Q \propto \delta_0^{-16/5}$,
hence $Q \propto h_o^{32/11}$.
Thus, the top and bottom border lines in FIG.~\ref{fig:bd1a} are
mapped to these curves.
Points on the left border line $\delta_0=0$ diverge in the physical plane,
and the remaining border $\delta_0=1/\xmax$ is mapped to a curve
close to the origin of the physical plane.
For each $N=1,2,\dots,8$ we have mapped the region in FIG.~\ref{fig:bd1a}
in this way, and the resulting physical parameter plane
is shown in FIG.~\ref{fig:physicalborder}.
Here, the typical set of parameter values in Appendix~\ref{appendixA} are used for physical
parameters other than $h_o$ and $Q$.

\subsection{Nonlinear model with $\Pi_3> 0$}\label{sec:Pi3Model}
An analysis of Eqs.~(\ref{s2}) and (\ref{s1}) where $\Pi_3>0$ is also possible, by numerical means, and partially also by using more intricate analytic arguments based on a Hamiltonian formalism which we only mention here for the sake of brevity.

To numerically solve the static equations including $\Pi_3>0$, a scheme similar to the one described above must be adopted. For this case, we have no analytic means of determining the period $T(C)$ in the same manner as above. However, one may integrate the equation system with an event solver: if we start integration from the corner, a half period is given when $\delta'=0$ (corresponding to the ``belly'' of the polygon), which defines the condition (event) to stop integration. While integrating from the corner $\delta_\textrm{\tiny min}$ to the belly $\delta_\textrm{\tiny max}$, one may also compute the period of the solution, $T=\tilde{T}(C)$, and so, invoke a secondary solver to solve for a $\delta_\textrm{\tiny min}$ such that the commensurability condition $T=\pi/N$ is satisfied. 
Such solutions have no extreme spikes as with $\Pi_3=0$ and  instead yield rounder corners. The periodicity $T$ of the solution does not any longer depend monotonically on the corner width of the roller, $\delta_\textrm{\tiny min}$, which results in a more complex solution space 
(some of the solutions look like the clover shapes reported by Bush \emph{et al}.~\cite{Aristoff}).
Interestingly, larger overlap of the parameter regions  corresponding to polygons with different symmetry number $N$ is seen, in particular for lower symmetries as low as $N\sim 2$. Details are discussed in~\cite{Martens,Watanabe}.

\section{Temporal  Stability of the Circular state}\label{sec:stab}

We now introduce time dependence to the equations.
First, we consider time-dependent version of the continuity equation (\ref{eq:masscons}).
\begin{eqnarray}
  \frac{\partial}{\partial t} \left[ \frac{R^2 - r_j^2}{2} \Delta h d \theta \right] & = &
  \left [ \frac{Q}{2\pi} d\theta + q_\theta(\theta) \right ]
\nonumber \\
  & - & \left [ \,q_r(\theta) d\theta + q_\theta(\theta+d\theta) \,\right ].
\label{masscons-dimensional}
\end{eqnarray}
The left hand side is an estimated rate of change in the volume of the roller slice.
The right hand side describes flux into and out of the volume as described
in Sec.~\ref{sec:massconservation}.
This equation becomes in dimensionless form:
\begin{eqnarray}
  {\cal T} \frac{\partial}{\partial t} \left[ \delta \left(1-\frac{\delta}{2}\right) \right] & = &
  \frac{1}{2\pi} - \xi_r(\theta) - \frac{\partial \xi_\theta}{\partial \theta}.
\label{eq:masscons-t}
\end{eqnarray}
Now, the roller thickness $\delta (t,\theta)$ also depends on $t$,
and the characteristic time scale ${\cal T}$ is
\begin{equation}
  {\cal T} \equiv {\frac{R^2 \Delta h}{Q}} \approx \frac{R^2 h_o^2}{Q} ,
\end{equation} 
i.e., time to fill the whole disk with radius $R$ and height $\Delta h \approx h_o$.
Using this together with the radial force balance equation (\ref{eq:radly_balance}) we get
\begin{equation}
\label{t2}
  {\cal T} \frac{\partial}{\partial t} \left[ \delta \left(1-\frac{\delta}{2}\right) \right] =
  \frac{1}{2\pi} - \frac{\Pi_1}{\delta} - \frac{\partial \xi_\theta}{\partial \theta} .
\end{equation}

Next, we introduce time dependence in the azimuthal momentum balance (\ref{s1}).
This becomes
\begin{eqnarray}
\nonumber
\frac{{\cal T}}{\Pi_4} {\frac{ \partial \xi_{\theta}}{ \partial t}} &=&
  -\frac{\partial \delta}{\partial \theta}
  -\left[ {\frac{\Pi_2}{\delta^2}}+ {\frac{\Pi_3 (2-\delta)}{2}}  \right] \xi_{\theta}\\
\label{t1}
&+&  3 \alpha {\rm Bo^{-2}} \delta \left[ {\frac{\rho_1'(\delta)}{\rho_1^2}} {\frac{ \partial \delta}{ \partial \theta}}+{\frac{\partial^3 \delta}{\partial \theta^3}}  \right] ,
\end{eqnarray}
where
\begin{eqnarray}
    \Pi_4 &\equiv&  \frac{gR^2 h_o^3}{Q^2}.
\end{eqnarray}

We now linearize around the circular solution as before: 
Setting $\delta = \delta_0 + \epsilon \delta_1(t,\theta)$ and 
$\xi_\theta = 0 + \epsilon \xi_1(t,\theta)$,
and taking the first order terms in $\epsilon$, we find
\begin{equation}
  {\cal T} (1-\delta_0) \frac{\partial \delta_1}{\partial t} =
  \frac{1}{2 \pi \delta_0} \delta_1 - {\frac{ \partial  \xi_1}{ \partial \theta}}
\end{equation}
and
\begin{eqnarray}
\nonumber 
  \frac{{\cal T}}{\Pi_4} {\frac{\partial \xi_1}{\partial t}} &=&
  - \left[ \frac{\Pi_2}{\delta_0^2} + \Pi_3 \left(1-\frac12\delta_0\right) \right] \xi_1\\
  & -& \left(1 - 3 \alpha B  {\rm Bo}^{-2 }\right) \frac{\partial \delta_1}{\partial \theta}
  + 3 \alpha \delta_0  {\rm Bo}^{-2}{\frac{\partial^3 \delta_1}{\partial \theta^3}}
\end{eqnarray}
instead of their stationary version (\ref{eq:LinAnalysis1a}) and (\ref{eq:LinAnalysis1b}).

We now Fourier transform by
\begin{equation}
  \left[ \begin{array}{c} 
    \delta_1(t,\theta) \\ \xi_1(t,\theta)
  \end{array} \right] = 
  \left[ \begin{array}{c} 
    z \\ x
  \end{array} \right] 
  \exp \left[ \frac{s t}{(1-\delta_0) {\cal T}} + ik \theta \right],
\end{equation}
where, due to the $2\pi$-periodicity in $\theta$, the wavenumber $k$ has to be an integer.
We find
\begin{eqnarray}
s x &=& -C_2 x -i(C_1 k+C_3 k^3) z  \\
s z &=&  - ik x + C_4 z,
\end{eqnarray}
where
\begin{eqnarray}
C_1 &=& (1-\delta_0) \Pi_4 \left(1 - 3 \alpha B {\rm Bo^{-2}} \right)\\
C_2 &=& (1-\delta_0) \Pi_4  \left[\frac{\Pi_2}{\delta_0^2}+\Pi_3\left(1-\frac12\delta_0\right)\right]\\
C_3 &=& \delta_0 (1-\delta_0) 3 \alpha \Pi_4 {\rm Bo^{-2}} \\
C_4 &=& \frac{1}{2 \pi \delta_0} ,
\label{eq:stabcoefficients}
\end{eqnarray}
where $B$ is defined in (\ref{B}).
The characteristic equation is
\begin{equation}
s^2 + (C_2-C_4) s+ (C_1 k^2+C_3 k^4)- C_2 C_4=0
\end{equation}
with solution
\begin{equation}
\label{spec1}
s = \frac12 \left[C_4-C_2 \pm \sqrt{(C_4+C_2)^2 -4(C_1k^2+C_3k^4)} \right].
\end{equation}

For $k=0$ the solutions are $s_+(0) = C_4$ and $s_-= -C_2$.
Since $C_4>0$, the spatially homogeneous (circular) mode is unstable, leading to an increase or decrease of $\delta$ away from $\delta_0$. This is clearly unphysical (since our circular state is uniquely defined from the outset) and reflects the fact that
flux conservation is not built into our time-dependent model, since (\ref{t2}) does not directly respect flux conservation.
Thus, we must require that the constraint (\ref{eq:totalmasscons2}) still holds,
whereby the total mass flux in and out of the roller is in balance.
To linear order, this condition implies that
\begin{equation}
\label{globcon}
\oint \delta_1 (\theta) d \theta = 0 .
\end{equation}
Thus, the mode $k=0$ must vanish and is excluded.
In the following we only consider integer values $k\geq 1$.

Let us take a look at the dispersion relations in (\ref{spec1}).
All the $\Pi$'s and the Bond number are positive, so all the $C$'s are guaranteed to be positive except $C_1$.
\begin{figure}[tbh]
    \includegraphics[width=0.6\hsize]{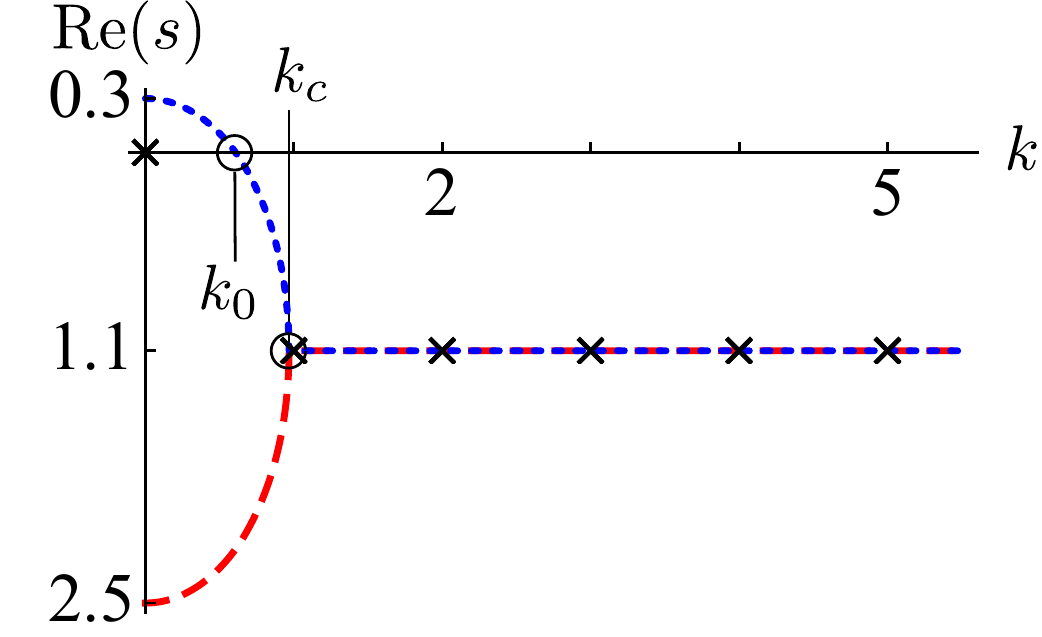}
    \includegraphics[width=0.6\hsize]{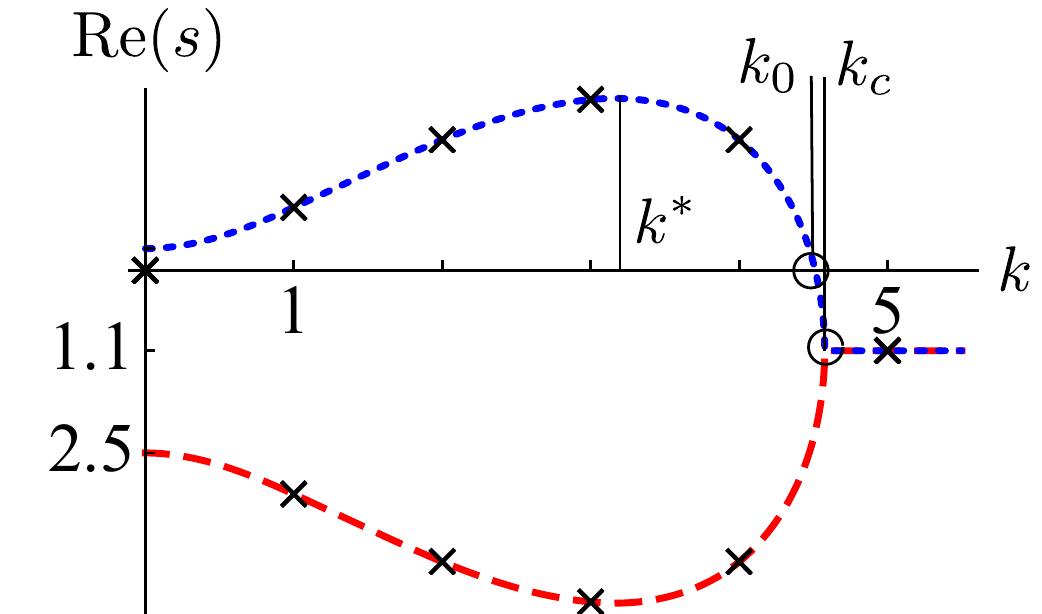}
    \caption{(Color online) 
	There is no maximally unstable wave number $k^*$ when $C_1>0$, see Eq.~(\ref{maxwavenumber}). The stability curves $s_+(k)$ (blue dotted) and $s_-(k)$ (red dashed) from (\ref{spec1}) for $C_1 = 2$ (above) and $C_1 = -2$ (below) with $C_2 = 2.5$, $C_3 = 0.1$ and $C_4 = 0.3$.     
        The dashed lines show the curves for continuous $k$, but only integral values of $k$, corresponding to the crosses, are allowed due to $2 \pi$-periodicity in $\theta$.
The mode $k=0$ is treated separately due to the global conservation law (\ref{globcon}).
\label{fig:disp1}}
\end{figure}
As discussed in Sec.~\ref{SurfaceTension}, surface shape parameter $B$ is positive when $R_1$ grows with $\delta$ -- this is the case at least close to the circular jump where $\delta\approx \delta_0$.
Then, the sign of $C_1$ only depends on the relative magnitude of $B$ and ${\rm Bo^2}$: 
when $3 \alpha B > {\rm Bo^2}$, we have $C_1<0$,
and surface tension is ``active'' in the sense that it then can cause a Rayleigh-Plateau-like instability.
This instability criterion is equivalent to 
\begin{equation}
\label{crit}
 \frac{R h_o }{l_c^2}< 3 B.
\end{equation}

At present, we do not have much data to check this criterion, but from the case presented in FIG.~\ref{fig:heightprofile} we can estimate the two curvature radii and thereby the magnitude of $B\approx 27$ (see Appendix A). 
Note however, that since the values entering $B$ should really be averages over the surface, we somewhat overestimate the curvatures and therefore likely underestimate $B$. 
Ethylene glycol has a surface tension of $\sigma \approx 50\times 10^{-3}$ N/m and a density of 1100 kg m$^{-3}$; these numbers yield a capillary length of $l_c \approx 2.2 $ mm and a Bond number Bo $\approx 2.3$. 
In effect, the left hand side of (\ref{crit}) thus becomes $\approx 32$, whereas the right hand side is $\approx 81$. This demonstrates that the circular state should indeed be unstable towards the formation of polygons. 
The flow-rate is $Q \approx 40$ ml s$^{-1}$ and
using the parameters listed in Appendix A, we have $\Pi_4 = gR^2 h_o^3 Q^{-2} \approx 0.69$; this yields a positive prefactor in $C_1$ of $(1-\delta_0) \Pi_4 \approx 0.4$.
Although the two sides of (\ref{crit}) appear similar, these estimates show that it is plausible that $C_1$ may take on negative values.

\begin{figure}[ht!]
    \includegraphics[width=0.6\hsize]{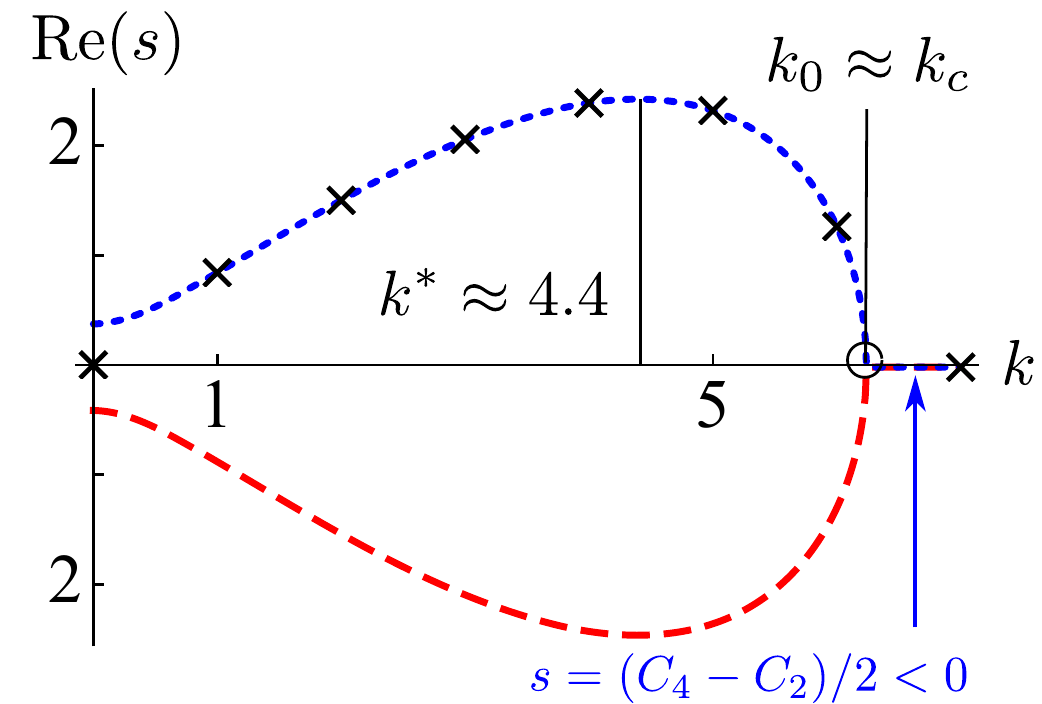}
    \caption{(Color online)  The stability curve $s_+(k)$ (dashed line) from Eq.~(\ref{spec1}) shown for stability coefficients $C_1 \approx -0.603$, $C_2 \approx 0.415$, $C_3 \approx 0.0156$, and $C_4 \approx 0.374$, based on experimental parameters and the height profile of a pentagon (see Appendix A). 
    Note that the constant branch $s(k>k_0) = (C_4-C_2)/2 \approx -0.02$ is negative.
    We then find that the maximally stable wave number (Eq.~\ref{maxwavenumber}) is $k^*\approx 4.4$; thus the most likely symmetry to occur has symmetry $k=4$ or $k=5$ which matches well the expectation.  The crosses mark the values of $s$ for integer $k$, respecting the $2 \pi$-periodicity. The mode $k=0$ is treated separately according to the global conservation law (\ref{globcon}). 
\label{fig:disp4}}
\end{figure}

Let us look at the dispersion relations in more detail.
First we note that $s(k)$ has reflective symmetry around $k=0$. In Figs.~\ref{fig:disp1}, \ref{fig:disp4} we therefore only display the positive $k$-axis. Along this positive $k$-axis, there are (at most) three distinct values that occur for the dispersion relation: (i) the maximal value of $s(k)$ at $k^*$, defining the two neighboring modes of strongest growth (integer values), (ii) the root of the dispersion relation $s(k_0)=0$, and (iii) the point $k_c$ where the two branches $s_-(k_c)=s_+(k_c)$ meet.
For $k>k_c$, the discriminant $D=(C_4+C_2)^2-4(C_1k^2+C_3k^4)$ becomes negative and the real part assumes simply a constant value, $\text{Re}(s)=(C_4-C_2)/2$, which defines a further symmetry axis for the dispersion relation. Note that, by definition, $k_0$ is identical to Eq.~(\ref{eq:klin}).
%

For the unstable case ($C_1 <0$) we expect the symmetry breaking predominantly to occur at  the \emph{maximally unstable wave number} $k^*$. $\text{Re}(s_+)$ is maximal when the term $|C_1|k^2 - C_3k^4$  is maximal, and so 
\begin{eqnarray}\label{maxwavenumber}
k^* = \sqrt{{\frac{-C_1}{2C_3}}}=\sqrt{\frac{3 \alpha B - \ {\rm Bo^2}}{6 \alpha \delta_0}}. 
\end{eqnarray}
Clearly, such a maximal value only exists when $C_1$ is negative, i.e.~when the condition $3\alpha B > \text{Bo}^2$ in ~(\ref{crit}) is satisfied, as illustrated in Fig.~\ref{fig:disp1}.
For small Bond number, or strong instability, (\ref{maxwavenumber}) simplifies to
\begin{eqnarray}
k^* \approx \sqrt{\frac{B}{2 \delta_0}} =\sqrt{\frac{ \rho_1'(\delta_0)}{2  \rho_1^2(\delta_0)}}.
\end{eqnarray}
and if the roller cross section is close to cylindrical, we find $k^* \sim \delta_0^{-1}$, i.e. a wavelength proportional to the width of the roller as in the Rayleigh-Plateau instability. This fits nicely with the observed bifurcation sequence, since, experimentally, polygons with many corners (up to 13) are seen close to the transition, where the roller is thin, and lower order polygons only occur later when the width is larger.

In FIG.~(\ref{fig:disp4}), we show the dispersion relation using parameter estimates from Appendix A which are based on the height profile of the pentagon shown in FIG.~\ref{fig:heightprofile}.
For this case, we obtain $k^* \approx 4.4$ and  $k_0\approx 6.23$. With this value of $k^*$, it is most likely that  polygons with either $k=4$ or $k=5$ corners occur, in accordance with the experimental observation.

\section{Discussion}\label{sec:discussion}

We hope that we have been able to convince the reader that  the type II hydraulic jumps and their intriguing polygon offspring can be understood within the phenomenological framework presented here. We have given a detailed presentation of the model and the underlying assumptions about the flow. In particular, we assume that the inner edge of a ``roller" structure inside the jump describes the shape of the type II jump. In the circular type II regime, the internal flow of this roller structure is similar to a toroidal vortex. 
The local width of this roller is one of three basic variables in our model. The two other variables in our model are the radial and tangential flow rates across the jump and inside the roller, respectively. A polygonal shape is obtained when the width loses circular symmetry and becomes a periodic function of the polar angle, while the outer edge of the roller remains circular, as observed in experiments. Also, for the polygonal states, a net flux is carried inside the roller from the sides to the corners. 
The model is formulated by considering a radial and a tangential force balance across the roller structure and the polygon shapes occur primarily by competition between viscous and gravitational effects. We have shown that one variant of the model is exactly solvable, even in the strongly non-linear limit, and that it has polygonal solutions which are quite similar to those observed experimentally. The resulting bifurcation diagram is qualitatively similar to experimental measurements.

The question of stability has been addressed as well. Here, surface tension plays an important role. For small deformations around the circular state, we 
can include surface tension
by considering the Laplace pressure inside the roller. This pressure can be estimated by introducing a parameter that characterizes how the curvature of the roller changes with deformation. When this parameter is positive, which is expected for the circular type II hydraulic jump, surface tension effects will tend to destabilize the circular state if the length scales at the jump (e.g. jump height $h_o$ and radius $R$) are of the order or smaller than the capillary length. Thus a polygon state will emerge with a wave length of the order of the width of the roller, in analogy with the Rayleigh-Plateau instability.

To include surface tension beyond the linear approximation would require more detailed knowledge about the height profile of the polygon states. Hopefully such results will soon be available --- indeed, one of the aims of the present paper is to stimulate renewed research in this direction.

In the present study, we have neglected the rotational motion inside the roller which is clearly observed in experiments. The stability of a rotating liquid column with free cylindrical surface has been investigated in other studies~\cite{Pedley,Kubitschek}. The relative importance of the rotational kinetic energy to the surface energy is characterized by the Weber number $\text{We}_{\Omega}=\rho a^3\Omega^2/\sigma$,
where $\Omega$ is the characteristic rotational frequency and $2a=R\,\delta$ is the width of the roller. 
For an infinitely long cylinder, subject to axisymmetric disturbances (neglecting disturbances in e.g. azimuthal direction), a necessary and sufficient condition for stability is given by $k^2 \geq 1+\text{We}_{\Omega}$~\cite{Pedley}, where $k$ is our wave number (this result is valid for potential vortex flows and solid body rotation), and thus rotation could widen the spectrum of unstable modes. 
Compared to the roller in the hydraulic jump, these vortices are of course strongly idealized, and we do not know at present how to include them in a convincing way.


\acknowledgments
We are very grateful to K{\aa}re Stokvad Hansen, Ditte J{\o}rgensen, Bj{\o}rn Tolbod, Jesper Larsen and Clive Ellegaard for valuable discussions in the initial phases of this work. We are particularly indebted to Johan R{\o}nby for valuable discussions and suggestions throughout the entire process. E.M. would also like to thank Steven Strogatz, Paul Steen, and Richard Rand for helpful discussions on issues regarding the stability analysis of strongly non-circular solutions.

\appendix
\section{Typical Parameters and Coefficients}
\label{appendixA}
A `typical' set of parameters used in the experiments is:
$Q \approx 40\times10^{-6}$ m$^3$/s,
$h_o \approx 5\times 10^{-3}$ m,
$h_i \approx 5\times 10^{-4}$ m,
$\nu \approx 10^{-5}$ m$^2$/s,
$g \approx 9.8$ m/s$^2$,
$\sigma \approx 50\times 10^{-3}$ N/m,
and 
$\rho \approx 1100$ \text{kg/m}$^3$.
Clearly, the assumption $h_i \ll h_o$ used throughout the article is then valid. With the above parameters, we have a Bond number of 
${\rm Bo} = 2.3$ (${\rm Bo^2} = 5.4$) and $l_c\approx2.2$ mm.

Furthermore, using the values from the height profile in Fig.~\ref{fig:heightprofile}, we read off the following numbers: the outer edge of the roller $R \approx 3\times 10^{-2}$~m; the width of the roller between two corners (belly) $R \delta_b=14$~mm, and in the corner $R\delta_c = 11.5$~mm; and the corresponding curvature radii $R_b=4.5$~mm and $R_c=2.5$~mm. Using the approximations $\delta_0=(\delta_b + \delta_c)/2\approx 0.426$ and
\begin{eqnarray*}
 B&=& \delta_0 \frac{d}{d\delta}\left.\frac{1}{\rho_1(\delta)}\right|_{\delta_0}\\
  &\approx&\frac{1}{2} \frac{\delta_b + \delta_c}{\delta_b-\delta_c}\left(\frac{1}{\rho_1(\delta_c)}-\frac{1}{\rho_1(\delta_b)}\right),\
\end{eqnarray*}
we estimate that $B\approx27$. Note that the values entering $B$ should really be averages over the surface; therefore we somewhat overestimate the curvatures and therefore likely underestimate $B$.

The scaling factors $c_1,c_2,c_3$ and $c_4$ are unknown, but using the above experimental parameters and the estimates from the height profile, we can at least estimate some of them.
Using the scaling $R  = c_4 Q^{5/8} \nu^{-3/8} g^{-1/8}$ proposed in~\cite{Dimon} for the outer edge of the roller, $R$, we estimate that $c_4\sim 0.3$; moreover, we can match the definition $\delta_0=2\pi \Pi_1$ with $c_1\sim 1.13$. The constants $c_2,c_3$ originate from the linearization of the frictional force (\ref{eq:dF_t_mu}), and it is reasonable to assume that they are of order unity, too.
With these estimates, the non-dimensional numbers become:
$\Pi_1 \approx 0.0678$,
$\Pi_2 \approx 0.00363$,
$\Pi_3 \approx 1.306$,
$\Pi_4 \approx 0.689$,
$\delta_0 \approx 0.426$,
$\phi \approx 11.6$.
The stability coefficients are then 
$C_1 \approx -0.603$,
$C_2 \approx 0.415$,
$C_3 \approx 0.0156$,
and
$C_4 \approx 0.374$.
With these estimated coefficients, we find the spectrum for stable wave numbers as shown in Fig.~\ref{fig:disp4}, with a maximally unstable wave number $k^*\approx 4.4$ and $s_+$ has a root $k_0\approx 6.23$. Evaluating the wave number of the linearly perturbed circular jump in Eq.~(\ref{eq:klin}), we have $k\approx 6.23$ which is at least quite close to a pentagon. However, we note again that these numbers can only serve as rough estimates due to the lack of experimental data (i.e. to determine $c_1,c_2,c_3,c_4$ and $B$).

Typical frequencies for the rotational motion in our experiment are of the order of 5 Hz, so $\Omega \sim 5 \times 2 \pi $ rad/s.
The roller structure has a width of order $R\delta_0/2 \approx 6$ mm and a height of $h_o/2=2.5$ mm. Using for the average roller radius an estimate of $(a=h_o+R\delta_0)/4\approx 4.5$ mm we arrive at an estimated Weber number of  $\text{We}_{\Omega}=\rho a^3 \Omega^2 / \sigma \sim 1.9$; notice though that the Weber number scales with the cube of the average roller width $a$ and is thus very sensitive to errors in the estimate of $a$. It is not clear that the vorticity extends all the way out to where the surface attains its maximal value (this is how we estimated $R\delta_0$), and therefore we may be overestimating $a$.

We define standard dimensionless numbers as follows.
Defining the velocity before the jump as $u_i = \frac{Q}{2 \pi R h_i}$ we get the Reynolds number 
\begin{equation}
\label{Re}
{\rm Re} = \frac{u_i h_o}{\nu}= \frac{Q h_o}{2 \pi R h_i \nu} \approx 212,
\end{equation}
where we take $h_o$ as the characteristic scale, similar to $R \delta$. 
The Froude numbers are
\begin{equation}
\label{Fri }
{\rm Fr_i} = \frac{u_i}{\sqrt{g h_i}}= \frac{Q}{2 \pi R h_i \sqrt{g h_i}},
\end{equation}
so
\begin{equation}
\label{Fri2 }
{\rm Fr_i^2} = \frac{Q^2}{(2 \pi)^2 g R^2 h_i^3 } \approx 37.037.
\end{equation}
Similarly, we have the downstream Froude number
\begin{equation}
\label{Fro2 }
{\rm Fr_o^2} = \frac{Q^2}{(2 \pi)^2 g R^2 h_o^3 } \approx 0.037.
\end{equation}
These dimensionless numbers are evaluated using the above choice of parameter, with $R=30$ mm for the height profile in Fig.~\ref{fig:heightprofile}.
With these definitions, $\phi$ from Eq.~(\ref{eq:phirelation}) is
\begin{equation}
\label{phi }
\phi= \frac{1}{4 \sqrt{c_1^3 c_2}}\,\frac{h_i^2}{R h_o} \,{\rm Fr_i^{-2}}\,{\rm Fr_o^{-2}}\, {\rm Re^2}.
\end{equation}

\noindent
The Weber number is defined as
\begin{equation}
{\rm We} = \frac{\rho\, u_i^2}{\sigma/h_o} 
\end{equation}
because $h_o$ is similar to $R \delta$, which is similar to radii of curvature.
Thus
\begin{equation}
\label{We }
{\rm We} =\frac{\rho h_o Q^2}{(2 \pi)^2 R^2 h_i^2 \sigma} .
\end{equation}

\section{Nomenclature}
\label{appendixB}
The variables and dimensionless constants used throughout the paper
are summarized in Table~\ref{nomenclature}.

\begin{table*}[ht]
\setlength{\extrarowheight}{0.3mm}
\begin{tabular}{l|l}
\hline \hline
  {\it Symbol} & {\it Meaning}\\
\hline
  $\sigma$ & surface tension\\
  $\rho$ & fluid density\\
  $\mu$ & dynamic viscosity\\
  $\nu$ & kinematic viscosity\\
  $Q$ & volumetric flow rate from nozzle \\
\hline
  $(r,z, \theta)$ & cylindrical coordinates\\
  $\mathbf{u}=\mathbf{u}(r,\theta,z)$& flow velocity\\
  $h=h(r,z,\theta)$ & fluid surface height\\
  $q_r$ & flow rate in radial direction\\
  $q_{\theta}$ & flow rate in azimuthal direction\\
  $\xi_r \equiv q_r/Q$ & non-dimensional radial flow rate\\
  $\xi_{\theta} \equiv q_{\theta}/Q$ & non-dimensional azimuthal flow
  rate\\
  $r_j(\theta)$ &  jump radius\\
  $\delta(\theta)\equiv\frac{R-r_j(\theta)}{R}$ &
  non-dimensional jump width\\
\hline
  $h_i$ & inner fluid height\\
  $h_o$ & outer fluid height \\
  $\Delta h \equiv h_o-h_i$ & height difference across jump\\
  $\delta_0$ & non-dimensional (circular) jump width \\
  $R$ & outer radius of roller structure\\
  $\alpha\equiv h_o/R$ & aspect ratio of jump\\
\hline
  $dA_f$ & free surface area of control volume\\
& with infinitesimal angle $d\theta$ (inf. CV)\\
  $dA_b(\theta)$ & bottom area of inf. CV\\
  $dA_r$ & back side of inf. CV\\
  $A_t(\theta)$ & lateral area of inf. CV\\
  $c_1,c_2,c_3,c_4$ & geometrical constants \\
  $R_1$ & principal radius of curvature (r,z)-plane\\
  $R_2$ & principal radius of curvature ($\approx$ (r,$\theta)$-plane)\\
  $\rho_1\equiv R_1/R$ & non-dimensional radius of curvature\\
  $B = \delta_0 (d/d\delta (1/\rho_1))_{\delta=\delta_0}$ & surface shape parameter\\
\hline
  $dF_r$ & radial force per infinitesimal angle $d\theta$\\
  $dF_{\theta}$ & azimuthal force per infinitesimal angle $d\theta$\\
\hline
  $N$ & polygon symmetry number\\
  $k$ & wave number\\

\hline
  $\Pi_1$ & hydrostatic vs. shear force (radial)\\
  $\Pi_2,\Pi_3$ & shear vs. hydrostatic forces (azimuthal)\\
  $\Pi_4$ & \\
  $\Bo \equiv h_o/l_c$ & Bond number\\
  $l_c\equiv \sqrt{\sigma/(\rho g)}$ & capillary length\\

\hline
  $\mathcal{T}$ & filling time: disk of radius $R$ with height $\Delta
  h$\\
  $C_1,C_2,C_3,C_4$ & parameters used in stability analysis\\
\hline
 $x = \delta/\delta_0$ & jump width (solvable model)\\
 $y = 2\pi \xi_{\theta}$ & transverse flow rate\\
 $\phi\equiv (2 \pi)^2 \Pi_1^{3/2} \Pi_2^{-1/2}$ & ''mass`` $\phi^2$\\
 $C,V$ & energy, potential function of ''mass`` $\phi^2$\\
 $\Theta\equiv \theta/\phi$ & rescaled angle \\
\hline \hline
\end{tabular}
\caption{Variables and parameters used in the model.}
\label{nomenclature}
\end{table*}


\begin{thebibliography}{99}

\bibitem{Rayleigh}
Lord Rayleigh, {\em Proc. Roy. Soc. London}, Ser. A {\bf 90}, 324 (1914).

\bibitem{Watson}
E.~J. Watson, {\em  J. Fluid Mech.}, {\bf 20}, 481 (1964)

\bibitem{Olsson}
R.~G. Olsson, and E.~T. Turkdogan, {\em Nature}, {\bf 211}, 813 (1966)

\bibitem{Chow}
V.~T. Chow, Open Channel Hydraulics (McGraw Hill, NY, 1959)

\bibitem{Nature}
C. Ellegaard, A. Espe Hansen, A. Haaning, K. Hansen, A. Marcussen, T. Bohr, J. Lundbek Hansen, and S. Watanabe,
{\em Nature} {\bf 392}, 767 (1998)

\bibitem{Nonlinearity}
C. Ellegaard, A. Espe Hansen, A.~Haaning, K. Hansen, A. Marcussen, T. Bohr, J. Lundbek Hansen, and S. Watanabe,
{\em Nonlinearity} {\bf 12}, 1-7 (1999)

\bibitem{PhysicaB}
T.~Bohr, C.~Ellegaard, A.~Espe Hansen, A.~Haaning,
{\em Physica B} {\bf 228}, 1-10 (1996)


\bibitem{Bohr}
T. Bohr, V.~Putkaradze and S.~Watanabe {\em Phys. Rev. Lett.}, {\bf 79}, 1038-1041 (1997)

\bibitem{WPB}
S. Watanabe, V. Putkaradze and T. Bohr, {\em J. Fluid Mech.}, {\bf 480}, 233 (2003)

\bibitem{RUC}
K.~Stokvad Hansen, D.~J{\o}rgensen, J.~R{\o}nby Pedersen and B.~Tolbod,
``Polygonformede hydrauliske spring - Et modelleringsprojekt" (in Danish)
{\em IMFUFA tekst nr. 411} (Roskilde University 2002).

\bibitem{Bush}
J.~W.~M. Bush, J.~M. Aristoff and A.~E. Hosoi,
{\em J. Fluid Mech.}, {\bf 558}, 33-52 (2006)


\bibitem{Duclaux}
V.~Duclaux, C.~Clanet and D.~Qu\'er\'e, {\em J. Fluid Mech.}, {\bf 556}, 217-226 (2006).

\bibitem{Martens}
E.~A. Martens: ''Hydraulic Jump. A Model of the Polygon Regime'', Master's Thesis, ETH Z\"urich, Z\"urich  (2004)

\bibitem{Watanabe}
S. Watanabe, ``A constrained variational model for symmetry breaking of the circular hydraulic jump", unpublished note (2008).

\bibitem{Dimon}
T. Bohr, P. Dimon and V.  Putkaradze, {\em J. Fluid Mech.},  {\bf 254}, 635 (1993)

\bibitem{1999ma}
A. Marcussen: ''Det hydrauliske spring; Et eksperimentelt studie af polygoner og hastighedsprofiler.'' (''The hydraulic jump; An experimental study of polygons and velocity profiles''). Master's thesis, CATS, Niels Bohr Institute, Copenhagen (1999). In Danish.

\bibitem{Aristoff}
J.~M. Aristoff, J.~D. LeBlanc, A.~E. Hosoi, J.~W.~M. Bush
{\em Physics of Fluids}, {\bf 16:9}, S4 (2004)



\bibitem{Pedley}
T.~J. Pedley, {\em J. Fluid Mech.},  {\bf 30}, 127-147 (1967)

\bibitem{Kubitschek}
J.~P. Kubitschek and P.~D. Weidman {\em J. Fluid Mech.}, {\bf 572}, 261-286 (2006)

\bibitem{suppmaterial}
Supplemental Materials with a visualization of the flow inside the roller structure, a demonstration of the role of surfactant eliminating polygons are provided at
\ifpdf \href{http://eam.webhop.net/SI.php5}{ http://eam.webhop.net/SI.php5}\else  \verb| http://eam.webhop.net/SI.php5 |\fi.




%

\end{thebibliography}
\end{document}